\documentclass[aps,preprint,amsmath,amssymb,amsfonts,nofootinbib]{revtex4-1}

\usepackage{epsfig}
\usepackage{graphicx}
\usepackage{dcolumn}
\usepackage{bm}
\usepackage[titletoc]{appendix}

\usepackage{graphicx}
\usepackage{subfigure}
\usepackage{amsfonts}
\usepackage{amsgen}
\usepackage{amsbsy}
\usepackage{amsmath}
\usepackage{mathrsfs}
\usepackage{mathtools}
\usepackage{epstopdf}
\usepackage{caption}
\usepackage{tabularx}
\usepackage{tabu}
\usepackage{xcolor}
\usepackage{nicefrac}
\usepackage{enumitem}
\usepackage{gensymb}

\newcommand\ie{\textit{i.e.~}}

\newcommand\eg{\textit{e.g.~}}
\newcommand{\e}[1]{\ensuremath{\times 10^{#1}}}
\newcommand{\Ge}[1]{\textcolor{black}{#1}}

\begin{document}

\author{Zhouyang Ge$^1$}\email{zhoge@mech.kth.se}
\author{Hanna Holmgren$^2$}
\author{Martin Kronbichler$^3$}\email{kronbichler@lnm.mw.tum.de}
\author{Luca Brandt$^1$}\email{luca@mech.kth.se}
\author{Gunilla Kreiss$^2$}\email{gunilla.kreiss@it.uu.se}
\affiliation{$^1$ Linn\'{e} FLOW Centre and SeRC (Swedish e-Science Research Centre), KTH Mechanics, SE-100 44 Stockholm, Sweden}
\affiliation{$^2$ Division of Scientific Computing, Department of Information Technology, Uppsala University, Box 337, 751 05 Uppsala, Sweden}
\affiliation{$^3$ Institute for Computational Mechanics, Technical University of Munich, Boltzmannstr.\ 15, 85748 Garching b.\ M\"{u}nchen, Germany}

\begin{abstract}
  
\Ge{Motivated by the emerging applications of liquid-infused surfaces (LIS), we study the drag reduction and robustness of transverse flows over two-dimensional microcavities partially filled with an oily lubricant.
Using separate simulations at different scales, characteristic contact line velocities at the fluid-solid intersection are first extracted from nano-scale phase field simulations and then applied to micron-scale two-phase flows, thus introducing a multiscale numerical framework to model the interface displacement and deformation within the cavities. 
As we explore the various effects of the lubricant-to-outer-fluid viscosity ratio $\tilde{\mu}_2/\tilde{\mu}_1$, the capillary number Ca, the static contact angle $\theta_s$, and the filling fraction of the cavity $\delta$, we find that the effective slip is most sensitive to the parameter  $\delta$. The effects of $\tilde{\mu}_2/\tilde{\mu}_1$ and $\theta_s$ are generally intertwined, but weakened if $\delta < 1$.
Moreover, for an initial filling fraction $\delta =0.94$, our results show that the effective slip is nearly independent of the capillary number, when it is small. Further increasing Ca to about $0.01 \tilde{\mu}_1/\tilde{\mu}_2$, we identify a possible failure mode, associated with lubricants draining from the LIS, for $\tilde{\mu}_2/\tilde{\mu}_1 \lesssim 0.1$. Very viscous lubricants (\eg $\tilde{\mu}_2/\tilde{\mu}_1 >1$), on the other hand, are immune to such failure due to their generally larger contact line velocity.}

\end{abstract}

\title{Effective slip over partially filled microcavities and its possible failure}
\maketitle

\section{Introduction}

Advances in microfluidics and nanotechnology have boosted a rapid development of surface engineering in the last two decades. Among the different effects of micro-/nano-patterned surfaces, often inspired by observations in nature, one remarkable finding is that the introduction of micro-/nano-scale roughness on an otherwise smooth hydrophobic surface can sometimes significantly reduce the resistance to an external liquid flow. This slippery effect, due to entrapment of gas or vapor pockets under the surface asperities (superhydrophobic Cassie state), was first observed in the experiment of a water flow through a water-repellent pipe \cite{Watanabe}. Subsequently, a number of studies have demonstrated various levels of drag reduction \cite{Ou,Choi_Kim,Schaffel,Lee}, but also in some cases drag enhancement \cite{Steinberger,Karatay}. Despite the discrepancies in the literature, a common technological challenge for the application of superhydrophobic materials is their fragility \cite{Bocquet}. Under high pressures or external forces, such as turbulent fluctuation or phase change, the surface texture can be partially or fully impregnated by the outer fluid (Cassie-to-Wenzel transition), causing the system to lose the features it was designed for \cite{Gentili,Giacomello, Seo_etal_18}.

Liquid-infused surfaces (LIS) are an alternative when aiming for drag reduction. They are more robust against pressure-induced failure, while displaying the same useful properties as conventional gas-cushioned superhydrophobic surfaces \cite{Wexler}. Two recent experiments have demonstrated, using microfabricated oil-impregnated pillars and grooves separately, up to 16\% drag reduction in laminar flows \cite{Solomon} and up to 14\% drag reduction in turbulent flows \cite{Rosenberg}. In the case of the turbulent flow, the authors also tested superhydrophobic surfaces and measured approximately 10\% drag reduction \cite{Rosenberg}.
\Ge{The values cited above, obtained at small lubricant-to-external-fluid viscosity ratios, can eventually decrease to nearly zero as the lubricant becomes more viscous. However, hybrid designs have been devised to maintain the performance, see \eg a recent proof-of-concept study \cite{Hemeda}.}

Analytically, the slippage over a superhydrophobic or liquid-infused surface can be characterized by an effective slip length. Analogous to the definition of the Navier slip, the effective slip length is an \textit{averaged} quantity equal to the distance below the surface at which the velocity would extrapolate to zero (to be distinguished from the \textit{intrinsic} slip of molecular nature \cite{Gentili}). Extensive studies have been devoted to obtaining theoretical expressions of the effective slip for two-dimensional longitudinal or transverse grooves \cite{Lauga_Stone, Sbragalia_Prosperetti, Davis_Lauga, Ng_Wang, Schonecker, Nizkaya, Crowdy_long, Crowdy_tran}.
\Ge{Among these, \cite{Lauga_Stone, Sbragalia_Prosperetti, Davis_Lauga, Ng_Wang, Crowdy_tran} assume perfect slip along the liquid-gas interface, \cite{Ng_Wang, Schonecker, Nizkaya} assume flat menisci, while the meniscus deformation, if considered, is either small \cite{Crowdy_long, Crowdy_tran} or in the dilute limit (\ie the surface is mostly solid) \cite{Lauga_Stone}. Furthermore, for purpose of calculation, the shape of the interface is always assumed symmetric (\ie flat or circular) even under shear. This practically limits the application of the analytical results to the zero capillary limit, being the upper/lower bound of the drag reduction depending on the specific conditions.}

Understanding the dependence of the slip length on the imposed shear and the lubricant viscosity in more realistic conditions may require a numerical approach. 
\Ge{There are, as yet, surprisingly few fully resolved hydrodynamic simulations able to solve the details of the flow reducing the underlying assumptions. Most prior numerical studies still consider flat/circular menisci with zero subphase viscosity \cite{Davies_etal, Martell_Perot_Rothstein, Cheng_Teo_Khoo, Wang_Teo_Khoo,Teo_Khoo_14, Seo_etal_18}; however, they extend analytical solutions to more complex surface patterns or the finite-Reynolds-number regime. Flexible bubble shapes were first considered in \cite{finland} for a uniform gas mattress, and later for a non-uniform distribution \cite{finland2}. Using a two-phase Lattice Boltzmann method, \cite{finland, finland2} show that increasing the capillary number reduces the effective slip, even below zero (\ie more friction than a solid plate). Specifically, their nanobubbles protrude strongly into the flow and remain trapped in the pores. Indeed, for very large protrusion angles, negative slip is both observed experimentally \cite{Steinberger} and verified analytically \cite{Davis_Lauga}. On the other hand, when the protrusion angle is smaller and the bubbles are allowed to slide on the substrate, the phase field simulation of \cite{Gao_Peng} shows the opposite behavior: the effective slip is nearly shear-independent for relatively low capillary numbers, while it can increase dramatically if the capillary number is beyond some threshold. This threshold is not a single, universal value but depends on the spacing of the grooves and the initial filling of the gas; however,  the enhancement of the slip is clearly due to depinning of the liquid-gas-solid contact line. We note that the depinning process considered in \cite{Gao_Peng} might be an idealization, since realistic solid surfaces may not be smooth/chemically-homogeneous near the edge. Furthermore, both studies consider gas bubbles submerged in water under unrealistically large shear rates ($10^6 \sim 10^7 s^{-1}$)\footnote{In \cite{finland}, the shear rates were reported as $10^{-6} \sim 10^{-7} s^{-1}$. This must be a typo.}. Whether this is stable or can be physically realized without generating significant heat remains an open question.}
  


Here, we explore a slightly different flow configuration: planar shear flows over a micro-rough wall \textit{partially} impregnated by a lubricant fluid. Using the newly developed multiscale numerical framework in \cite{Hanna}, simulations at separate scales are performed to obtain the steady drag reduction, while capturing the dynamic wetting behavior in details.
\Ge{As we investigate the various effects of the viscosity ratio, the capillary number, and the static contact angle, we find that the filling fraction has the largest impact for drag reduction. It weakens the effects of other parameters, which are generally intertwined in a number of non-trivial ways. Moreover, for a given initial filling fraction (94\%), our results show that the viscosity of the lubricant can not only influence the effective slip length, but also the robustness of the substrate under external shear. Shear-driven failure of LIS has recently been reported in \cite{Wexler, Jacobi, Liu} in the longitudinal case. Our study predicts that a similar drainage, though the viscosity dependence differs, may also occur in the transverse case.}
Understanding of this drainage failure is instructive for improved robustness of the surface design.


\section{Microcavities partially filled with lubricants} \label{definition}

\subsection{Problem setup} \label{setup}

\begin{figure}[t]
 \begin{center}
 \includegraphics[width=\columnwidth]{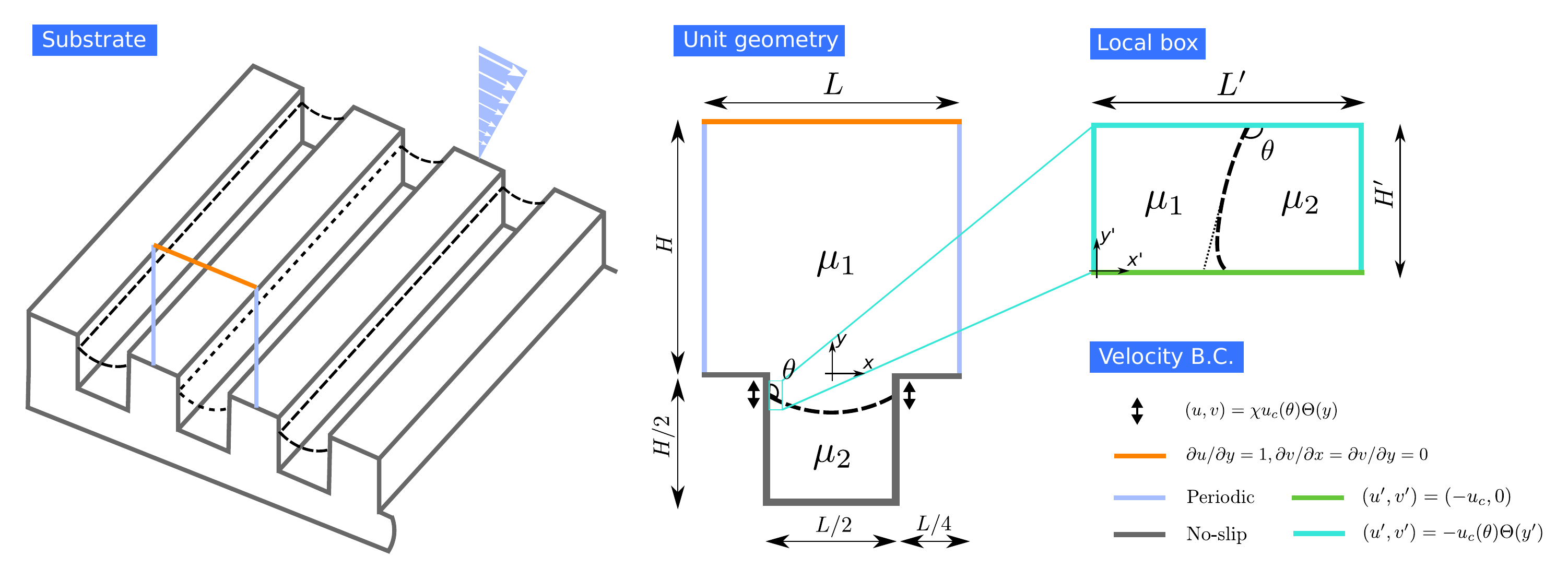}
 \end{center}
 \caption{(color online) Schematic of the problem definition and setup for the two separate simulations. The substrate is patterned with an array of square cavities. The unit geometry shows the cross-section of the partially filled microcavity. The boundary condition at $y=H$ is equivalent to unit tangential stress and zero normal stress, while the arrows near the contact lines represent the slip boundary. The local box depicts the computational domain of the moving contact line model (the variables are denoted by a prime). Its velocity boundary conditions correspond to a moving wall in the bending interface reference frame.}
 \label{fig: schematic}
\end{figure}

We consider the transverse flow over an array of regularly spaced square cavities illustrated in Fig.\ \ref{fig: schematic}. The outer fluid of viscosity $\mu_1$ is driven by a constant shear $\dot{\gamma}$ in the $x$ direction, imposed at distance $H$ above the floor. The cavities of length $L/2$ and depth $H/2$ are partially filled with a lubricant fluid of viscosity $\mu_2$. When the number of the microcavities is large, the system is equivalent to a single cavity with periodic boundary conditions in the front and back. The solution at the (quasi-)steady state is determined by the incompressible Stokes equations, written in the non-dimensional form
\begin{equation} \label{Stokes}
  \begin{aligned}
    \nabla \cdot {\bm u} = 0, \quad
    -\nabla p + \nabla \cdot \big[ \mu_i ( \nabla {\bm u} + \nabla {\bm u}^T ) \big] = 0,
  \end{aligned}
\end{equation}
where ${\bm u}=(u,v)$ is the velocity, $p=p(x,y)$ the pressure, and $\mu_i=\tilde{\mu}_i/\tilde{\mu}_1$ ($i=$ 1 or 2) the dimensionless viscosity, using $\tilde{H}$ and $\dot{\tilde{\gamma}} \tilde{H}$ as the reference length and velocity respectively\Ge{\footnote{Dimensional values are denoted with a tilde throughout the manuscript.}}. For viscous flows, the velocity and its tangential derivatives are continuous along the fluid interface \cite{Batchelor}. The normal stress is discontinuous due to the surface tension $\tilde{\sigma}$ and the viscosity difference, giving the pressure jump (denoted as $[A]_\Gamma=A_2-A_1$)
\begin{equation} \label{p jump}
  [p]_\Gamma = \frac{\kappa}{\textrm{Ca}} + 2[\mu]_\Gamma {\bm n}^T \cdot \nabla {\bm u} \cdot {\bm n} 
  \quad \textrm{on} \quad \Gamma,
\end{equation}
where ${\bm n}$ is the outward-pointing normal at the interface $\Gamma$, $\kappa$ its curvature, and Ca $=\tilde{\mu}_1 \dot{\tilde{\gamma}} \tilde{H}/\tilde{\sigma}$ the capillary number. 

\Ge{As the lubricant only partially fills the cavity initially, it may become distorted or splatter under the external shear. The associated contact line motion can be described by a second capillary number,}
Ca$_c=\tilde{\mu}_2 \tilde{U}_c/\tilde{\sigma}$, where $\tilde{U}_c$ is the characteristic contact line velocity related to the liquid and solid surface energies. The ratio between this velocity and the shear, $\chi=\tilde{U}_c/ (\dot{\tilde{\gamma}} \tilde{H})$, measures the magnitude of the local slip in the sheared dynamical system. It also scales the slip velocity near the contact line,
\begin{equation} \label{slip bc}
  \begin{aligned}
    {\bm u} = \chi u_c (\theta) {\bm \Theta} (y) \quad \textrm{on} \quad \partial \Omega_\Gamma,
  \end{aligned}
\end{equation}
where $\chi u_c(\theta)$ is the renormalized nanoscale contact line velocity depending on the apparent contact angle $\theta$, and ${\bm \Theta}(y)$ provides the self-similar slip velocity function of the wall-parallel coordinate $y$ that is imposed in the vicinity of the contact line on the boundary $\partial \Omega_\Gamma$ (see Fig.\ \ref{fig: schematic}). Further details of $\chi u_c(\theta)$ and ${\bm \Theta}(y)$ will be provided in Sec.\ \ref{model}.

In summary, Eqs.\ (\ref{Stokes}--\ref{slip bc}) are determined, neglecting the fluid inertia and fixing the substrate geometry, by the following non-dimensional parameters: 
\textit{(i)} the viscosity ratio, $\tilde{\mu}_2/\tilde{\mu}_1$, 
\textit{(ii)} the static contact angle, $\theta_s$, 
\textit{(iii)} the \Ge{initial filling fraction} of the cavity \Ge{$\delta=2d_0/H$} (where \Ge{$d_0$ is the initial depth of the lubricant measured from the contact point to the bottom of the cavity}),
\textit{(iv)} the capillary number based on the imposed shear Ca, and
\textit{(v)} the ratio between the characteristic contact line velocity and the shear, $\chi$.
The effect of the presence of the lubricating cavity and the corresponding apparent slip can be readily quantified by an effective slip length $\lambda_e$, defined as
\begin{equation} \label{eff slip}
  \lambda_e = \frac{\bar{u}(H)}{\dot{\gamma}H}-1,
\end{equation}
with $\bar{u}(H)$ being the average streamwise velocity at distance $H$ above the floor (averaged over the $x$ direction).

\Ge{In the following, we will consider various combinations of the governing parameters \textit{(i--v)} and evaluate $\lambda_e$ for each configuration. As the result will clearly depend on the motion of the impregnated lubricant, the multiscale modelling approach that we adopt is described next. The objective here is to provide an overall description of our methodology, rather than deriving the full mathematical/numerical details. For the latter, including validations, we refer to our previous work \cite{Martin, Hanna}.}

\subsection{Modelling of the moving contact lines} \label{model}

\Ge{We model the contact line dynamics in two steps.} First, we solve the Cahn-Hilliard equations within a Stokes system
\begin{equation} \label{C-H}
  \frac{\partial c}{\partial t} + \tilde{{\bm u}}\cdot \nabla c -\tilde{m}\nabla ^2 \tilde{\psi}=0,
  \quad \tilde{\psi} - \frac{3\tilde{\sigma} \tilde{\epsilon}}{4} 
  \bigg(\frac{2}{\tilde{\epsilon}^2}(c^3-c) -\nabla ^2 c \bigg) =0,
\end{equation}
\begin{equation} \label{C-H Stokes}
    \nabla \cdot \tilde{{\bm u}} = 0, \quad
    -\nabla \tilde{p} + \nabla \cdot 
    \big[ \tilde{\mu} ( \nabla \tilde{{\bm u}} + \nabla \tilde{{\bm u}}^T ) \big] +\tilde{\psi}\nabla c = 0.
\end{equation}
In the above, $c$ is a non-dimensional phase parameter smoothly varying from $+1$ in one fluid to $-1$ in the other within a thickness of $\tilde{\epsilon}$, $\tilde{\psi}$ is the fluid chemical potential, $\tilde{m}$ is the mobility, and $\tilde{\sigma}$, again, is the surface tension. The chemical potential $\tilde{\psi}$ measures the variation of the system free energy with respect to $c$. Its gradient determines the interfacial diffusion flux $-\tilde{m}\nabla \tilde{\psi}$, which together with the convective flux $\tilde{{\bm u}} c$, models the creation, movement, and dissolution of phase interfaces \cite{Jacqmin2000}. 

Technically, Eqs.\ (\ref{C-H}--\ref{C-H Stokes}) are solved in a rectangular box in the vicinity of a contact line using methods presented in \cite{Martin} (see Fig.\ \ref{fig: schematic}, the local box and its velocity boundary condition). 
\Ge{They are determined solely by the viscosity ratio, the surface tension, and the static contact angle (the rest are fixed choosing the proper non-dimensionalization);} hence, the moving contact line can be simulated separately from the cavity flow.
Inherently, we assume the length and time scales of the local box are much smaller than the cavity, \ie $\tilde{H}'/\tilde{H} \ll 1$ and $\tilde{\tau}'/\tilde{\tau} \ll 1$ respectively. The first condition holds by definition and is enforced by providing enough resolution. The second condition is automatically satisfied realizing $\tilde{\tau}'/\tilde{\tau} =\tilde{H}'/\tilde{U}_c/\dot{\tilde{\gamma}}^{-1}=\tilde{H}'/\tilde{H} /\chi$. We will show in Sec.\ \ref{results} that $\chi$ in our case is indeed much bigger than 1.


\Ge{The steady-state solutions of Eqs.\ (\ref{C-H}--\ref{C-H Stokes}) give the contact line velocity, $\chi u_c(\theta)$, function of the apparent contact angle only. It is typically nonlinear, and is valid down to the nanometer scale. To impose this slip velocity in the micrometer cavity flow, as the second step, we modify the velocity boundary condition near the contact line using asymptotic matching \cite{Hanna}. Here, the self-similarity of the local velocity field is invoked and the singularity of the viscous stress is avoided \cite{Huh_Scriven}. The end result is an algebraic operator, ${\bm \Theta}(y)$, applied to $\chi u_c(\theta)$ on the boundary $\partial \Omega_\Gamma$.}

\Ge{
We comment that our multiscale modelling approach is not limited to the phase-field model for the nanoscale; in principle, any model able to describe the contact line dynamics, \eg the molecular dynamics (MD) \cite{Johansson} or the Lattice-Boltzmann (LB) \cite{Sbragaglia_etal}, can be used. We also note that, by solving Eqs.\ (\ref{C-H}--\ref{C-H Stokes}) in a square domain, we implicitly assume the solid surface is nanosmooth.}
Consequently, any deviation from the static contact angle will result in an interface displacement, bringing the phase field back to its local equilibrium. In practice, a real surface may have random roughness or defects smaller than the scale of the printed patterns, causing the interface to be pinned (\ie contact angle hysteresis). Such effects can be included by modifying the geometry of the computational domain, or simply by modifying the relation $u_c=u_c(\theta)$ so that $u_c=0$ for a range of $\theta$'s. 
\Ge{In Sec.\ \ref{results}, we will take this second approach to account for a small contact angle hysteresis.}

\subsection{Numerical methods} \label{numm}

The governing equations, together with the boundary conditions, Eqs.\ (\ref{Stokes}--\ref{slip bc}), are solved numerically using the two-phase flow solver described in \cite{MartinHPC}, with suitable modifications for moving contact lines. The equations are discretized in space using the finite element method and the solver is implemented in the C++ based finite element open source library \texttt{deal.II} \cite{DEAL1, DEAL2}.
\Ge{The interface between the two fluids is evolved using the conservative level set method \cite{CONSLS}, so that only one fixed set of mesh is required. Specifically, we use uniformly distributed quadrilaterals (\ie squares) with grid spacing $\Delta x=1/160$, and time steps restricted by the stability condition $\Delta t_{max} = c_0$Ca$\Delta x$ ($c_0$ is a constant) \cite{MartinHPC}. This leads to $\Delta t =10^{-4} \sim 10^{-3}$ depending on the capillary number Ca. }

\Ge{The moving contact-line velocities are pre-computed by solving Eqs.\ (\ref{C-H}--\ref{C-H Stokes}) and used as tabulated inputs.
In the simulations, additional numerical parameters include the frequency of the reinitialization (a technical procedure in the level set method, see \cite{CONSLS}), the size of the local box in the contact line model (\ie $L'$ and $H'$, see Fig.\ \ref{fig: schematic}), and the size of a so-called bump function (related to $\partial \Omega_\Gamma$, see \cite{Hanna}). These are chosen to yield numerically-independent results 
as in  \cite{Martin, Hanna} where validations are presented.}

\section{Results} \label{results}

We study the effective slip over microcavities partially filled with a second fluid using the parameters summarized in Tab.\ \ref{tab: param}.
Here, six pairs of fluids are considered as in the experiment \cite{Solomon}, leading to a wide range of $\tilde{\mu}_2/\tilde{\mu}_1$ from $31.7$ to $3.83\e{-3}$. \Ge{The filling fraction is initialized to $\delta=0.94$ (corresponding to a depth $d_0=0.47H$) to allow for some sloshing of the lubricant.} The velocity ratio $\chi=\tilde{U}_c/ (\dot{\tilde{\gamma}} \tilde{H})$ is set constant for all the fluid pairs and shear rates to reduce the number of parameters and focus on the single physical effects mentioned above. This also implies that the capillary number is varied by changing the outer fluid viscosity. As an example, for $\dot{\tilde{\gamma}}=800$ s$^{-1}$, $\tilde{H}=20$ $\mu$m, and $\tilde{U}_c=2.63$ m/s (see \cite{dimension} for the detailed estimation), the velocity ratio $\chi \approx 164$ (which is indeed much greater than 1), and the corresponding Ca increases from $1.92\e{-3}$ to $1.59$ for the different viscosities considered. We further modify Ca at a fixed $\chi$ to study the effect of interface deformation. Finally, the effect of the static contact angle is investigated by considering $\theta_s=80\degree$ (leading to a convex meniscus), $\theta_s=105\degree$ (concave meniscus), \Ge{and $\chi u_c(\theta)=0 $ for $\theta_s \in [76\degree, 84\degree ]$ to mimic some contact angle hysteresis.}

\begin{table}[t]
  \centering
  \caption{Parameters for the outer (subscript 1) and lubricant (subscript 2) fluids in the present study.}
  \bgroup
  \def\arraystretch{1.05} 
  {\setlength{\tabcolsep}{0.9em} 
  \begin{tabular}{l l l c c c c}
    \hline
    $\tilde{\mu}_1 [\nicefrac{kg}{ms}]$ 
    &$\tilde{\mu}_2 [\nicefrac{kg}{ms}]$ 
    &$\nicefrac{\tilde{\mu}_2}{\tilde{\mu}_1}$ 
    &$\delta$
    &$\chi$
    &Ca
    &$\theta_s$ (deg)\\
    \hline
    0.0024 &0.0760 &$31.7$        &0.94  &164 &$0.02 \sim 5$  &80 or 105 or $76 \sim 84$\\
    0.0024 &0.0076 &$3.17$        &0.94  &164 &$0.02 \sim 5$  &80 or 105 or $76 \sim 84$\\
    0.0152 &0.0076 &$0.5$         &0.94  &164 &$0.02 \sim 5$  &80 or 105 or $76 \sim 84$\\
    0.1504 &0.0076 &$5.05\e{-2}$  &0.94  &164 &$0.02 \sim 5$  &80 or 105 or $76 \sim 84$\\
    0.8942 &0.0076 &$8.50\e{-3}$  &0.94  &164 &$0.02 \sim 5$  &80 or 105 or $76 \sim 84$\\
    1.9850 &0.0076 &$3.83\e{-3}$  &0.94  &164 &$0.02 \sim 5$  &80 or 105 or $76 \sim 84$\\
    \hline
  \end{tabular}}
  \egroup
  \label{tab: param}
\end{table}

\subsection{Motions at the contact line}

\begin{figure}[t]
 \begin{center}
   \includegraphics[width=.9\columnwidth]{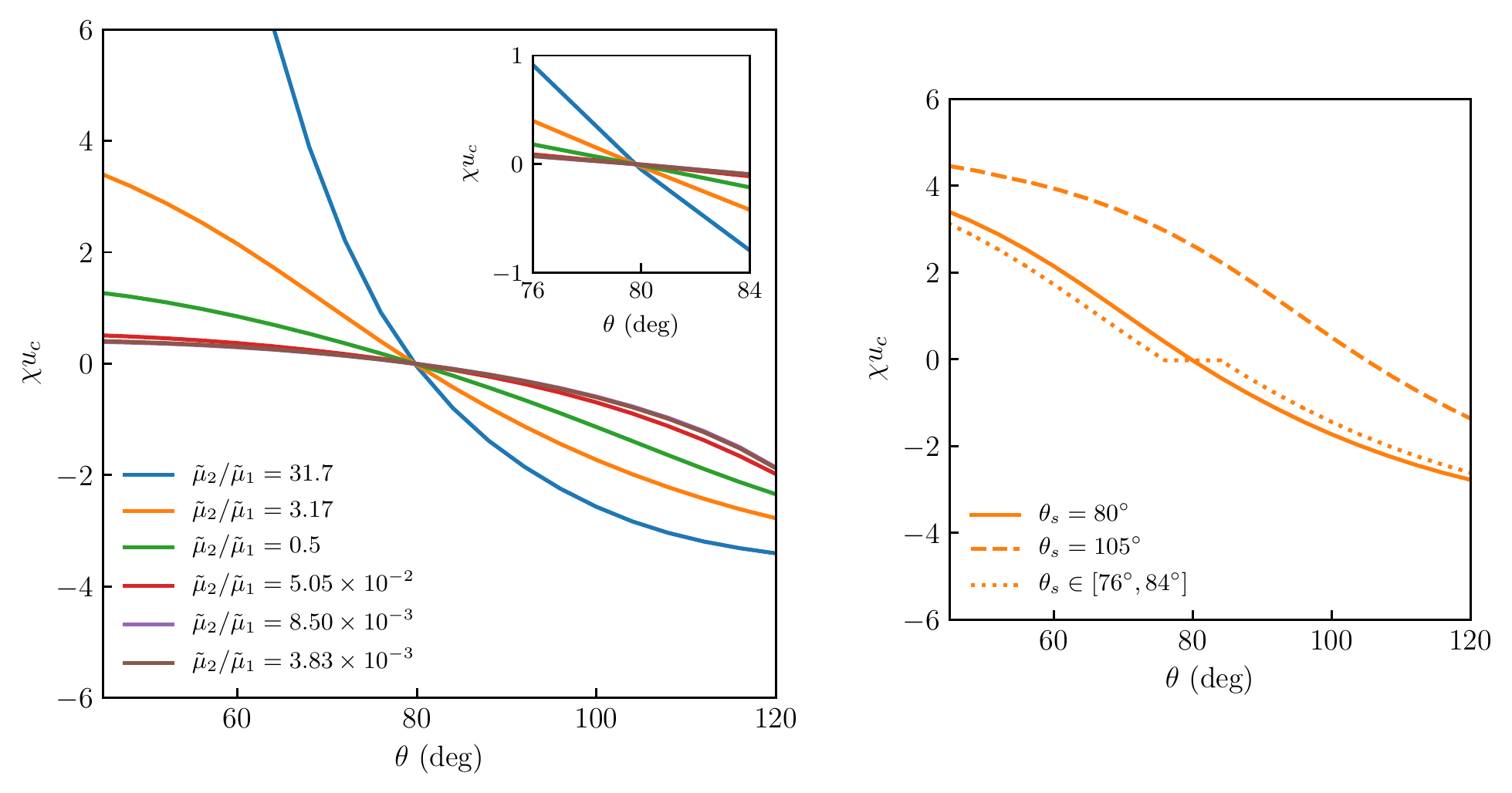}
 \end{center}
 \caption{(color online) Relations between the apparent contact angles and the contact line velocities for $\theta_s=80\degree$ under various viscosity ratios, precomputed using the contact line model described in Sec.\ \ref{model}. \Ge{Inset shows a close-up at small angle deviations, whereas the panel on the right illustrates how we model different static angles and contact angle hysteresis.}}
 \label{fig:tabulated}
\end{figure}

We precompute the contact line velocity $\chi u_c$ as function of contact angle $\theta$ for the range of parameters listed in Tab.\ \ref{tab: param}, using the nanoscale phase-field model described in Sec.\ \ref{model}.
Numerically, the non-dimensional height of the local box is $H'=36$, with grid size $h = 36/128$ and time step $\Delta t=0.5$. Steady state results obtained after $4000$ time steps are plotted in Fig.\ \ref{fig:tabulated}. 
Here, the solid lines correspond to static contact angle $\theta_s=80\degree$ measured from the outer fluid side. The non-zero contact line velocity at $\theta \neq \theta_s$ shows the tendency of the contact line to reach its equilibrium position. For the results presented next, we have also used static contact angles $\theta_s=105\degree$, corresponding to menisci protruding into the cavity, and $\theta_s \in[76\degree , 84\degree]$, modelling a contact angle hysteresis of $8\degree$. Keeping the rest of the parameters unchanged, the contact line velocities for  $\theta_s=105\degree$ and the case with hysteresis are obtained by shifting the curves pertaining each viscosity ratio horizontally to the modified static angles, see Fig.\ \ref{fig:tabulated} (right) for an example.  

\Ge{As we vary $\tilde{\mu}_2/\tilde{\mu}_1$ over four orders-of-magnitude, Fig.\ \ref{fig:tabulated} reveals a non-trivial dependence of the contact line dynamics. On the one end, $\chi u_c$ changes rapidly with $\theta$ for very viscous lubricants, almost diverging for $\theta < 60\degree$ in the case of $\tilde{\mu}_2/\tilde{\mu}_1=31.7$; on the other end, as the lubricant becomes less and less viscous, the $\chi u_c (\theta)$ relations eventually collapse onto one curve. Qualitatively, reduction of $\chi u_c$ for decreasing $\tilde{\mu}_2/\tilde{\mu}_1$ is expected as we normalize the flow using the shear in the outer fluid; in other words, it is easier (hence requires less velocity) to displace a less viscous fluid (\ie deviating $\theta$ from $\theta_s$). In addition, the change of the curvature of the function $\chi u_c ( \theta)$  can be inferred from the reciprocity of the two fluids (\ie $-\chi u_c$ instead of $\chi u_c$ and $ 180-\theta$ instead of $\theta$ for the same $\tilde{\mu}_1/\tilde{\mu}_2$)\footnote{This is merely a qualitative argument, as it does not preserve the static angle unless $\theta_s=90\degree$.}. Quantitatively, the present model has been compared favorably with Cox' law \cite{COX}, especially for small angle deviations \cite{Martin}. Since this is the regime where the fluids normally operate at, we expect our model to accurately capture the small-scale contact line motions.}

\Ge{Finally, we note that the slope of the contact line velocity profiles near the static contact angle, $\theta_s$, (cf.\ Fig.\ \ref{fig:tabulated} inset) plays an important role in the wetting of the cavity under external shear.} As we will discuss later, the difference of the contact line velocity with the viscosity ratio completely alters the robustness of lubricant infused cavities.



\subsection{Effective slip above the cavities} \label{eslip}

\begin{figure}[t]
 \begin{center}

 \includegraphics[width=.6\columnwidth]{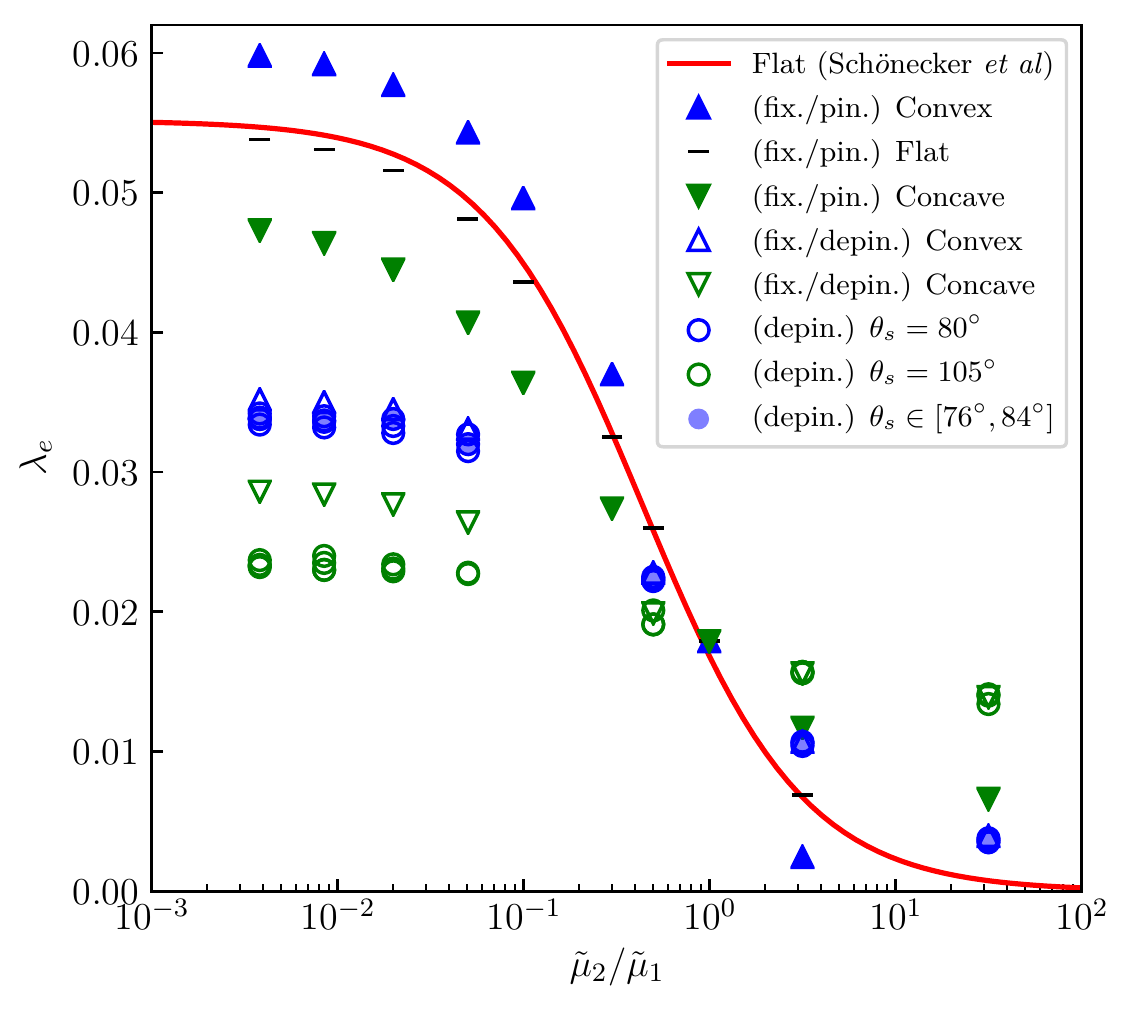}

 \begin{picture}(0,0)
   \put(-80,70){\includegraphics[height=1.8cm]{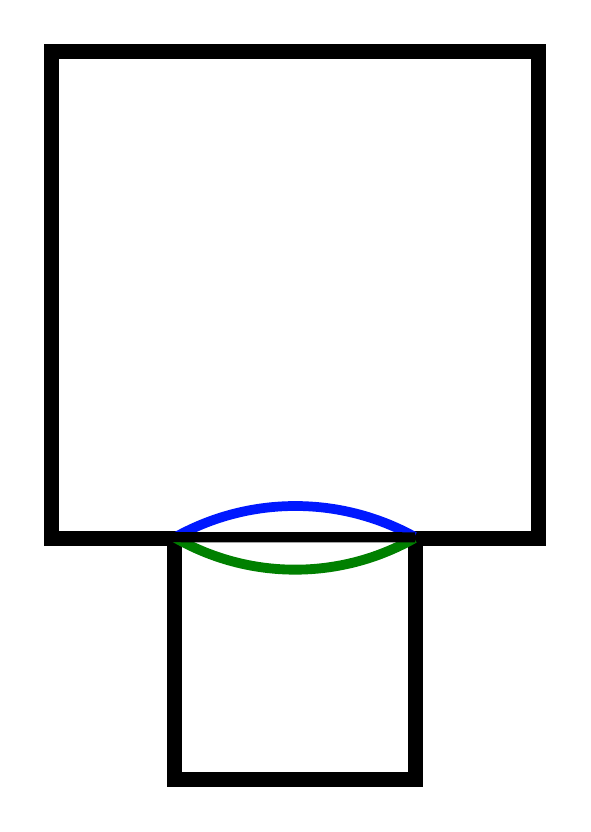}}
   \put(-30,70){\includegraphics[height=1.8cm]{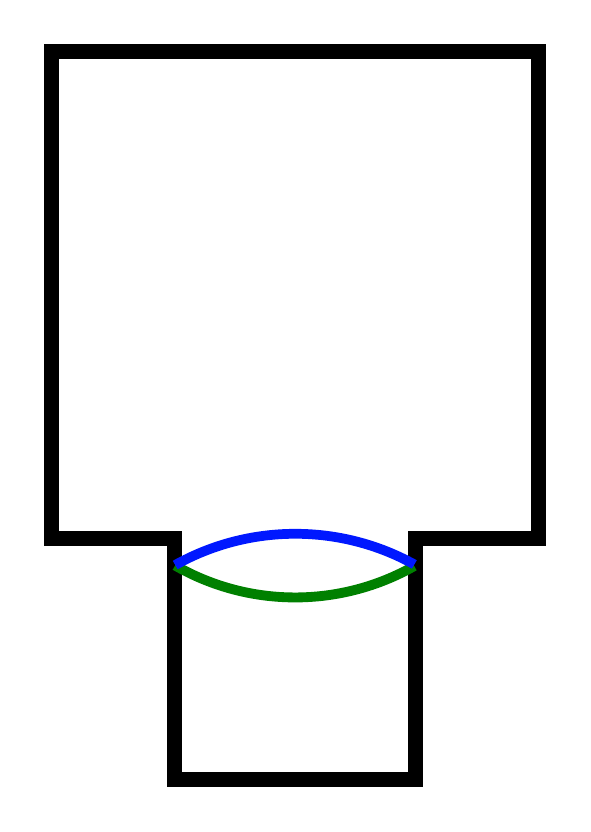}}
   \put(-80,62){fix./pin.} 
   \put(-30,62){ depin.} 
 \end{picture}
 \end{center}
 \caption{(color online) Effective slip as function of viscosity ratio under various static contact angles, filling fractions, and capillary numbers. The bars represent flat and fixed interfaces fully covering the cavity ($\delta=1$), where the analytical result from \cite{Schonecker} is also plotted (red line). The filled symbols, upper blue and lower green triangles, stand for convex ($\theta_s=80\degree$) or concave ($\theta_s=105\degree$) interfaces pinned at the cavity tip in the zero capillary limit (\ie fixed interface). The open symbols are the steady state solutions at $\delta=0.94$ for different $\theta_s$ and Ca. \Ge{The capillary number Ca is not indicated as it does not affect the results noticeably.}}
 \label{fig: slip-visc}
\end{figure}

Now, we present steady-state results of the effective slip length, defined in Eq.\ \eqref{eff slip}, obtained by solving the governing Eqs.\ (\ref{Stokes}--\ref{slip bc}) for the setup depicted in Fig.\ \ref{fig: schematic}, using the two-phase Stokes solver described in Sec.\ \ref{numm}. \Ge{The overall results are compiled and plotted in Fig.\ \ref{fig: slip-visc}, divided into the following two categories. First, we discuss the results obtained fixing the interface shape and pinning the contact point at the cavity corner, and compare with existing theories (denoted as ``fix./pin."). In a later section, we present results with depinned interface, \ie contact line not at the cavity corner, obtained both fixing the interface (``fix./depin.") and letting it move according to the multiscale model presented above (``depin.").}

\subsubsection{Fixed interfaces pinned at the corners.}
  
\Ge{Partly as a validation of our numerical methods, we first consider interface of fixed shapes pinned at the cavity tips.}
These are obtained by imposing in the simulation flat/circular menisci fully covering the cavities, indicated by bars/filled triangles in Fig.\ \ref{fig: slip-visc}. Comparing with the analytical model taking into account finite dissipation in the cavity \cite{Schonecker}, a close agreement is observed over the broad $\tilde{\mu}_2/\tilde{\mu}_1$ spectrum examined. 
Specifically, the results show a continuous decrease of the effective slip as the viscosity ratio increases; the rate of variation is logarithmic for $0.1 < \tilde{\mu}_2/\tilde{\mu}_1 < 10$, it begins to saturate for $\tilde{\mu}_2/\tilde{\mu}_1 \lesssim 0.1$, and it is practically zero for $\tilde{\mu}_2/\tilde{\mu}_1 > 10$. 

We further examine the curvature dependence of the effective slip length, using $\theta_s = 80\degree$ and $105\degree$ as two representative curvatures for weakly convex and concave interfaces respectively.
As shown in Fig.\ \ref{fig: slip-visc},
for $\tilde{\mu}_2/\tilde{\mu}_1 < 1$, weakly convex interfaces have larger slip than weakly concave ones, consistently with previous analytical and experimental studies \cite{Davis_Lauga, Karatay}. 
The difference of the effective slip between convex and concave menisci increases in the limit of zero viscosity ratio; this is 
approximately 25\% bigger in the convex  case.
\Ge{On the other hand, when $\tilde{\mu}_2/\tilde{\mu}_1 > 1$, the relative magnitude flips: the concave interfaces have a larger slip length than the flat ones, while the convex interfaces can even have negative slip, adding more drag to the flow.} 
The reason for this asymmetry is rather straightforward. 
Similar to the reasoning in \cite{Sbragalia_Prosperetti}, the increased shear stress modified by a more viscous fluid will reduce the local slip, even more so when the interface bows into the channel, hence a smaller $\lambda_e$ for the convex meniscus than for the concave one.

\Ge{Lastly, we remark that the dependence of the effective slip length on the curvature is non-trivial. Previous studies have shown the existence of a critical contact angle beyond which the effective slip becomes negative ($\theta_s \lesssim 30\degree$ by our definition) \cite{Davis_Lauga, Karatay, finland}. The angles we consider here are far from that range.}

\subsubsection{Interfaces depinned from the corners.}

\Ge{Next, we allow the interface to deform and slide on the cavity walls under external shear, removing the constraint of edge pinning considered earlier. The data pertain the steady state configuration, reached for shorter times at smaller Ca and verified to be unaffected by any numerical perturbations. Specifically, the effective slip length obtained initializing the filling ratio of the cavity to $\delta=0.94$ are displayed with open or round symbols in Fig.\ \ref{fig: slip-visc}.}

\begin{figure}[t]
 \begin{center}
 \includegraphics[width=.6\columnwidth]{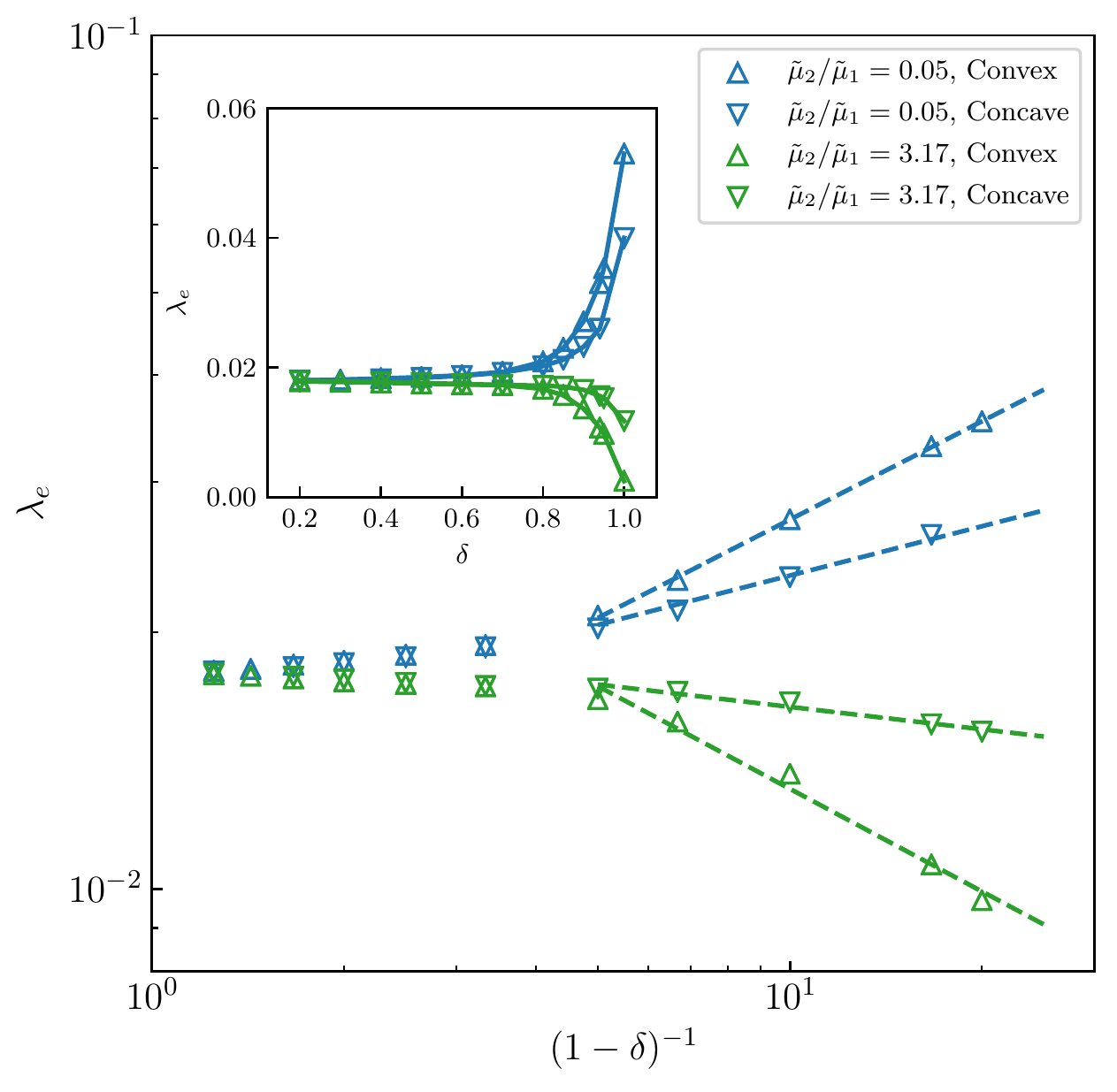}
 \end{center}
 \caption{Effective slip of partially filled cavities for convex ($80\degree$) and concave ($105\degree$) interfaces for $\tilde{\mu}_2/\tilde{\mu}_1=0.05$ (blue triangles) and 3.17 (green triangles) in the zero capillary limit. The main figure shows power law relations of the slip length when plotted against the inverse ``void fraction'' $(1-\delta)^{-1}$, indicated by the dashed lines (linear least squares fits for $(1-\delta)^{-1}>5$, or equivalently $\delta > 0.8$). The inset shows the sharp reduction/increase of the slip length as the meniscus recedes.}
 \label{fig: slip-height}
\end{figure}

\Ge{First, we note that the effective slip length of partially filled cavities differs appreciably from the fully covered ones, regardless of the contact angle and the capillary number. For very low ($\tilde{\mu}_2/\tilde{\mu}_1 < 0.1$) and very high ($\tilde{\mu}_2/\tilde{\mu}_1 > 10$) viscosity ratios, the difference is at least a factor of 2. Meanwhile, within the cases considered for depinned interfaces, the effect of the viscosity ratio on the slip length is weaker than it is for pinned interfaces. The overall variation of $\lambda_e$ is reduced.}
These observations suggest that the filling fraction of the cavity may be the main factor determining the effective slip. 

To examine possible relationships between $\lambda_e$ and $\delta$, we display in Fig.\ \ref{fig: slip-height} the effective slip under various filling fractions, for both convex ($\theta_s=80\degree$) and concave ($\theta_s=105\degree$) interfaces, at $\tilde{\mu}_2/\tilde{\mu}_1=0.05$ and 3.17. For this extensive parameter study, fixed interface shapes are imposed, corresponding to the zero capillary limit (minimum-energy interface) to speed up the simulations. Indeed, having a small capillary number does not affect the result, as shown in Fig.\ \ref{fig: slip-visc} where the slip length for the ``fix./depin." cases are not significantly different from the depinned cases at low Ca.

\Ge{As shown in Fig.\ \ref{fig: slip-height}, the effective slip length clearly depends on the filling fraction: as the meniscus recedes from the cavity tip, $\lambda_e$ quickly decreases or increases depending on $\tilde{\mu}_2/\tilde{\mu}_1$; the variation is the sharpest in the early stage ($0.8 \lesssim \delta < 1$), while it is nearly negligible as $\delta$ further reduces.
Plotting $\lambda_e$ against $(1-\delta)^{-1}$, which may be interpreted as an inverse ``void fraction'' of the cavity, we find a power law relation between the effective slip and the filling fraction. Indicated by the dashed lines in Fig.\ \ref{fig: slip-height}, the effective slip behaves as $\lambda_e \sim (1-\delta)^{-c}$ for $\delta > 0.8$, where $c$ is a constant related to the viscosity ratio and the overall geometry. Specifically, using linear least squares, we find $c \approx 0.38$ (convex) and $\approx 0.19$ (concave) for $\tilde{\mu}_2/\tilde{\mu}_1=0.05$, and $c \approx -0.40$ (convex) and $\approx -0.09$ (concave) for $\tilde{\mu}_2/\tilde{\mu}_1=3.17$. At equal viscosities, the $\lambda_e$--$(1-\delta)^{-1}$ relations display cone-like patterns with the spreading angle function of both the viscosity ratio and the meniscus curvature. At lower filling ratios, all the points converge to the value of the slip length of the single-phase cavity. 
We remark that a theoretical determination of c is likely difficult, as the governing equation here is biharmonic \cite{Crowdy_tran, Ng_Wang}.}
 Nevertheless, our results clearly illustrate the pronounced dependence of the effective slip on the interface displacement, already when small, for transverse grooves. 

\Ge{We note from above that the slip $\lambda_e$ varies in opposite directions depending on $\tilde{\mu}_2/\tilde{\mu}_1$. This is also shown in Fig.\ \ref{fig: slip-visc}, where the effective slip of the depinned interfaces intercepts the red line (\ie the results for a flat interface) at $\tilde{\mu}_2/\tilde{\mu}_1 =1$ for both contact angles under consideration. \Ge{Specifically, the effective slip $\lambda_e$ is the same for $\delta=0.94$ and $\delta=1$, if $\tilde{\mu}_2 = \tilde{\mu}_1$;} when $\tilde{\mu}_2 < \tilde{\mu}_1$, \Ge{the slip is larger for $\delta=1$; when $\tilde{\mu}_2 > \tilde{\mu}_1$, on the contrary, it is larger for} $\delta=0.94$. This crossover thus suggests an additional viscosity dependence of the effective slip coupled with the filling fraction of the cavities.} 

\begin{figure}[t]
 \begin{center}
 \subfigure[\quad Flat (tip)]
 {\includegraphics[width=.42\columnwidth]{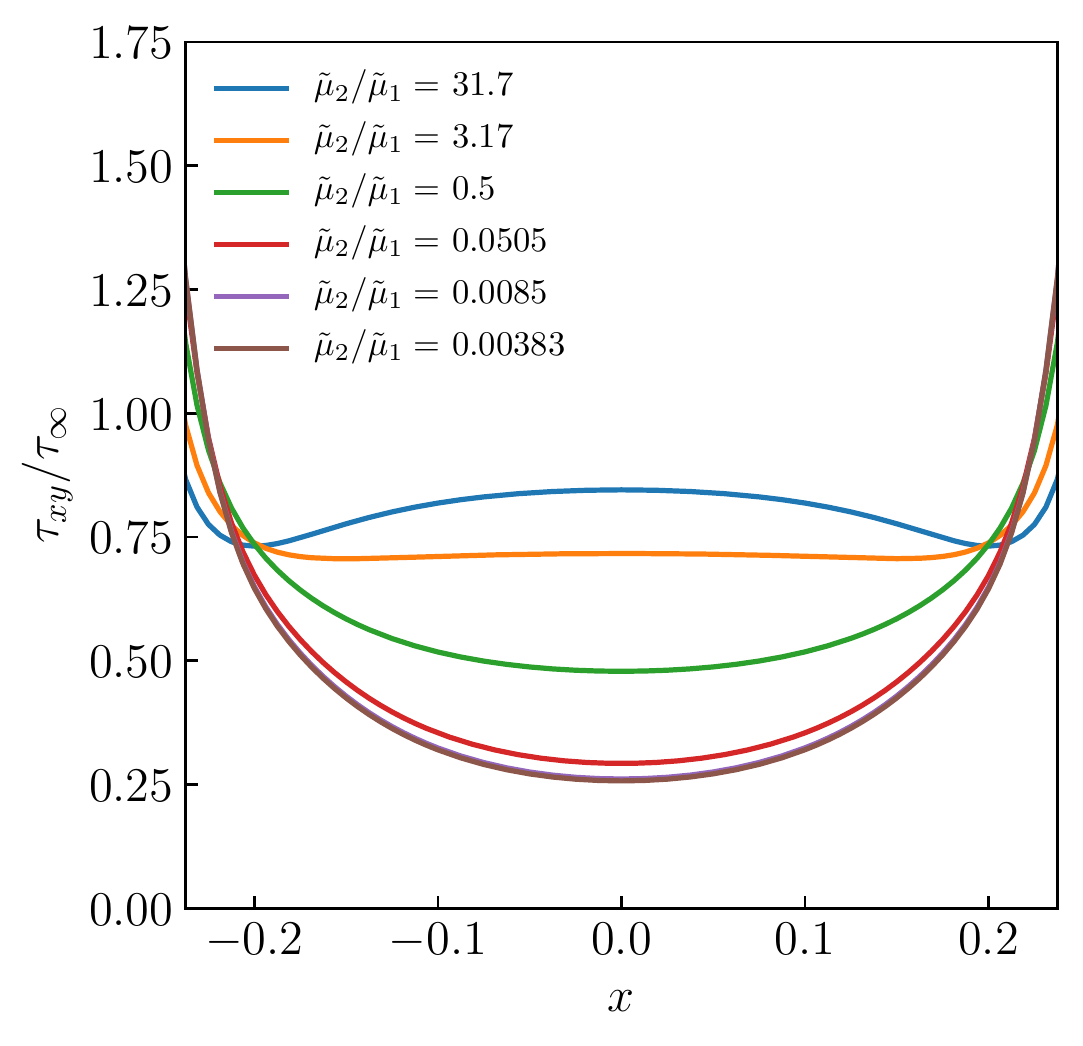}}
 \subfigure[\quad Flat (meniscus)]
 {\includegraphics[width=.42\columnwidth]{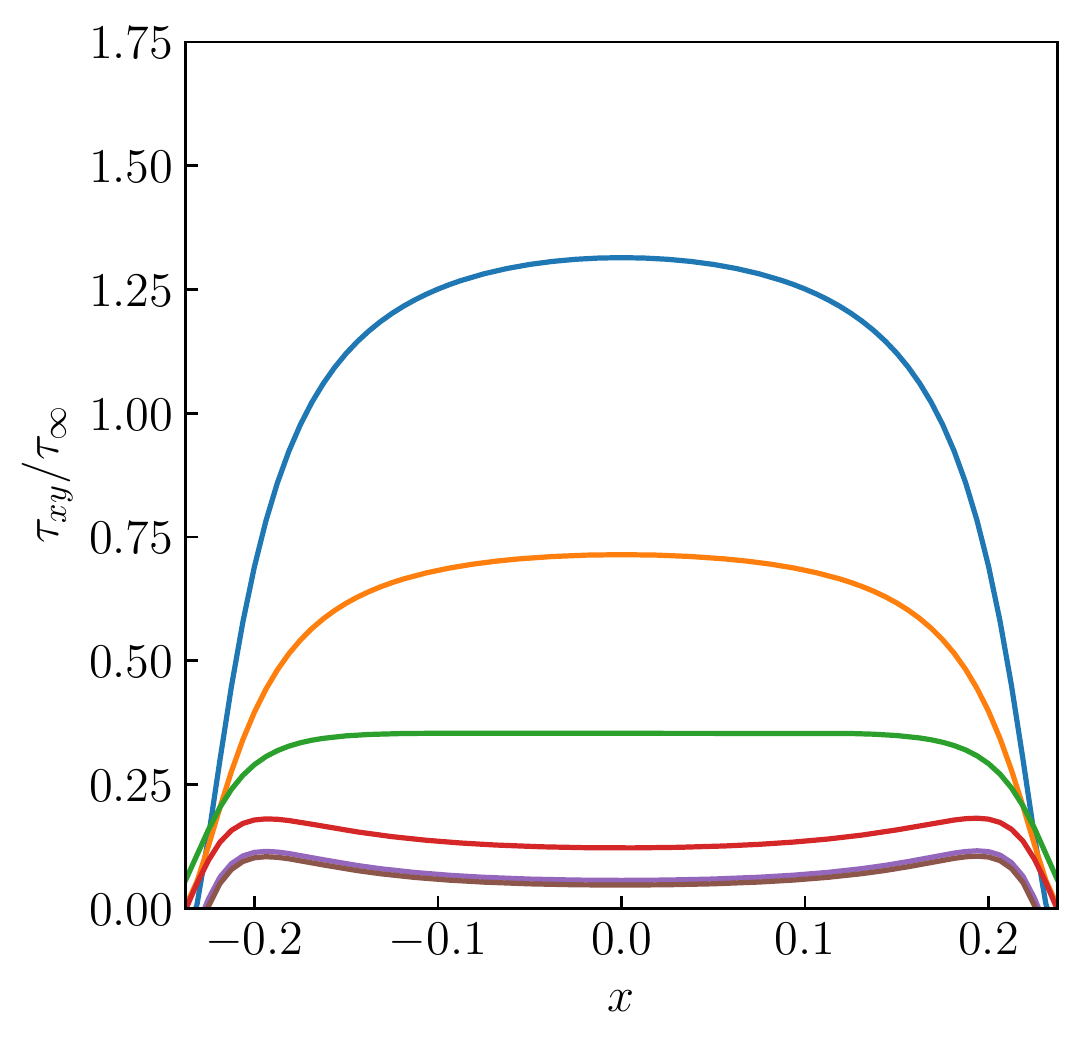}}
 \subfigure[\quad Convex (tip)]
 {\includegraphics[width=.42\columnwidth]{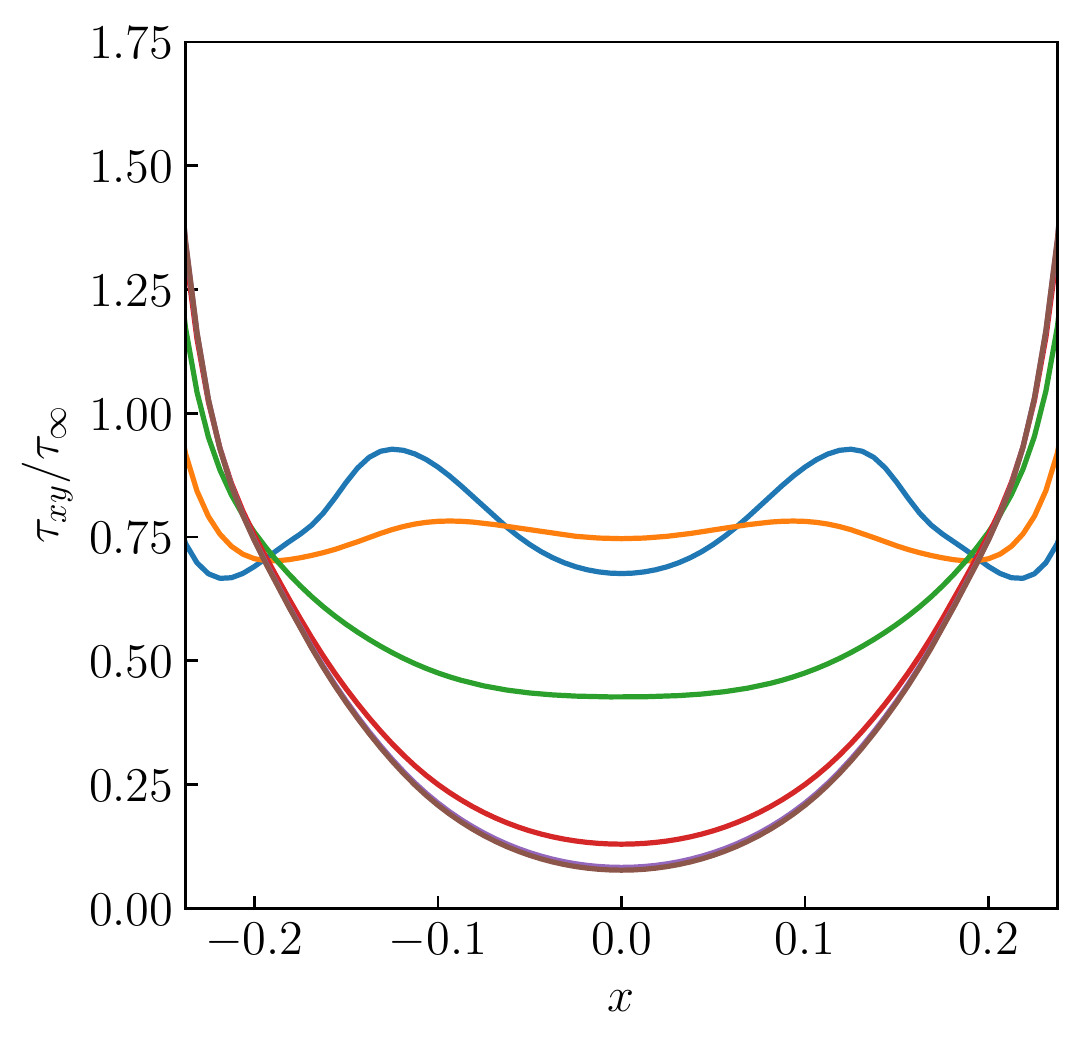}}
 \subfigure[\quad Concave (tip)]
 {\includegraphics[width=.42\columnwidth]{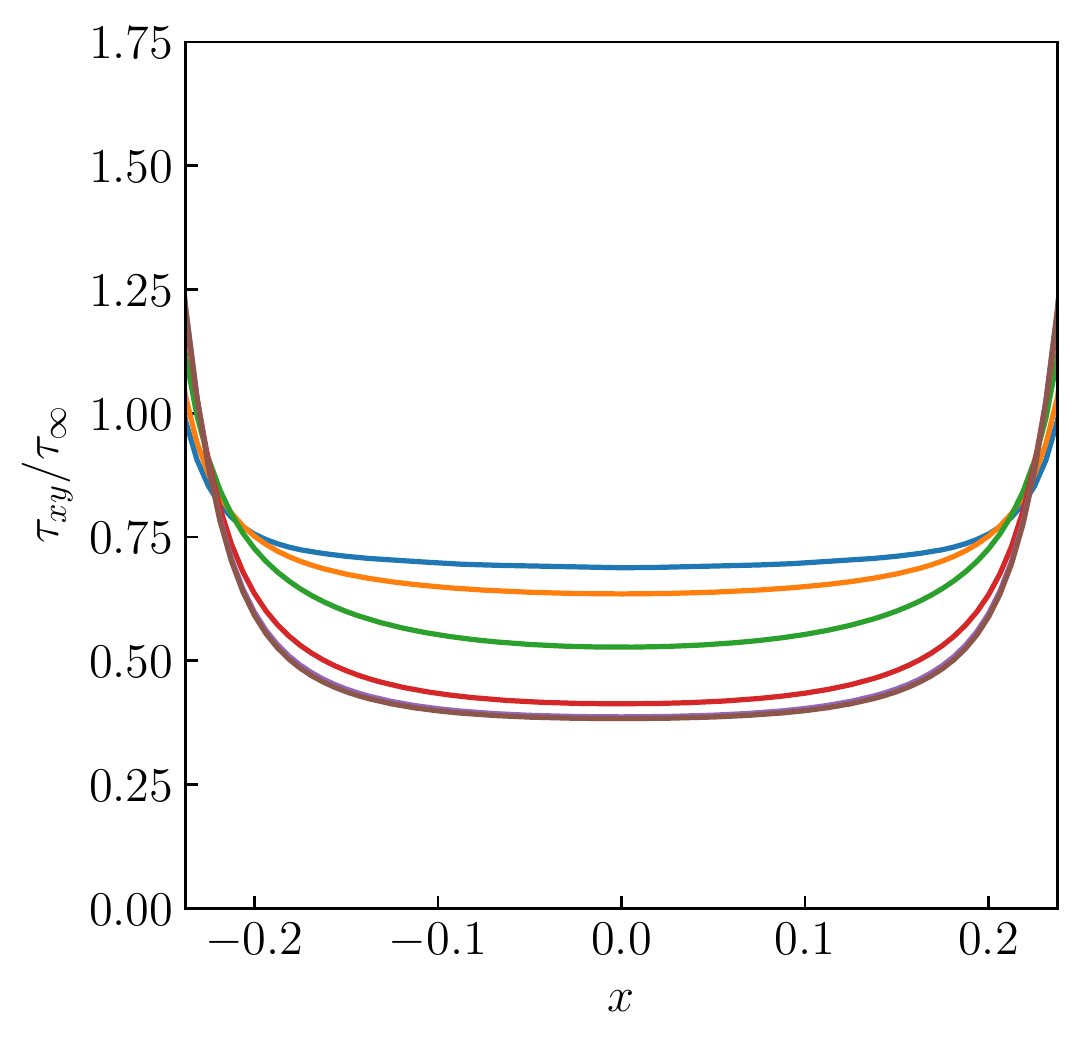}}
 \end{center}
 \caption{Normalized tangential stress evaluated on the plane of the cavity tip (a, c, d) or along the menisci (b) for various viscosity ratios. In all the cases, the menisci are fixed and depinned from the corners, corresponding to filling fraction $\delta=0.94$.}
 \label{fig: stress}
\end{figure}

\Ge{To quantitatively compare the effect of the viscosity ratio on $\lambda_e$ at $\delta=0.94$, we display the tangential shear stress $\tau_{xy}$ for flat, convex, and concave menisci in Fig.\ \ref{fig: stress}. Here, $\tau_{xy}$ is evaluated either on the cavity tip (at $y=0$) or along the fluid-fluid interface (at $y=-0.03H$), and it is normalized by the unit tangential shear stress $\tau_\infty$ imposed above the floor (at $y=H$). 
As shown in Fig.\ \ref{fig: stress}, the normalized shear stress decreases as we reduce $\tilde{\mu}_2/\tilde{\mu}_1$ for all the cases, consistent with enhanced slip at lower lubricant viscosities; however, $\tau_{xy}/\tau_\infty$ does not converge to zero as it would have been if the cavities were fully covered.
Close comparison of Fig.\ \ref{fig: stress}(c) and (d) also explains the flipping of the relative magnitude of $\lambda_e$ between convex and concave interfaces noted above: the shear stress is less for convex interfaces when $\tilde{\mu}_2/\tilde{\mu}_1 <1$, while it is less (on average) for the concave ones at $\tilde{\mu}_2/\tilde{\mu}_1 >1$. 
Moreover, Fig.\ \ref{fig: stress} reveals that the distribution of the local shear for partially filled cavities is non-uniform. When $\tilde{\mu}_2/\tilde{\mu}_1 <1$, $\tau_{xy}/\tau_\infty$ always retains its minimum value at $x=0$, and increases gradually towards the walls (at $x=\pm 0.25$); when $\tilde{\mu}_2/\tilde{\mu}_1 > 1$, the shear stress profiles can have several local minima/maxima depending on the protrusion angle. Such non-uniformity is most prominent when the interface is convex. In general, both the viscosity of the two fluids and the geometry of the liquid-infused cavities appear to influence $\tau_{xy}/\tau_\infty$.}

\Ge{We remark that constant shear stress along substrate surfaces is sometimes assumed in theoretical models to obtain analytical  solutions \cite{Schonecker}. Although it is verified for fully-covered flat cavities (see Fig.\ 4 in \cite{Schonecker}), our results suggest that it is inaccurate for partially filled ones, even along the fluid-fluid interface, see Fig.\ \ref{fig: stress}(b). Since liquid-infused substrates are not always fully-covered in practice \cite{Wexler}, our simulations suggest this assumption be relaxed when developing more comprehensive models.}

Finally, we discuss the role of capillary and hysteresis, referring back to the circular symbols in Fig.\ \ref{fig: slip-visc}. These data are obtained via the multiscale contact line model for capillary numbers Ca $=0.02\sim 5$ and contact angles $\theta_s=80\degree$, $105\degree$, or $76\degree \sim 84\degree$ at initial filling fraction $\delta=0.94$. 
\Ge{Surprisingly, we find virtually no influence of the contact angle hysteresis on the effective slip length for the entire range of viscosity ratios considered. The filled circles, corresponding to $\theta_s \in [76\degree, 84\degree$], lie closely on top of the blue open circles denoting $\theta_s=80\degree$. 
Our results thus provide evidence that small scale roughness on the substrate surface, due to the material itself or the fabrication precision, does not necessarily increase the  overall drag over the cavities. Indeed, as discussed in \cite{Schonecker}, the effective slip length is a far-field effect determined by the mean velocity above the substrate. Since a small contact angle hysteresis does not alter significantly the interface profile nor its wetting behavior, these changes are expected to be quickly smeared out away from the substrate. } 

The above reasoning applies only to the cases when the capillary number is small. 
Further increasing the capillary number, hence the shear, can eventually deform the fluid interface to an extent that a stable configuration may not be attainable. In the remaining, we will consider the lubricant-infused surface under extreme shear rates. As we examine the possible consequence under various conditions, a seemingly counter-intuitive technical solution will be suggested.

\subsection{Possible drainage of the lubricant}

\begin{figure}[t]
  \centering
  \subfigure[$\quad \theta_s=80\degree$]
  {\includegraphics[width=.35\columnwidth]{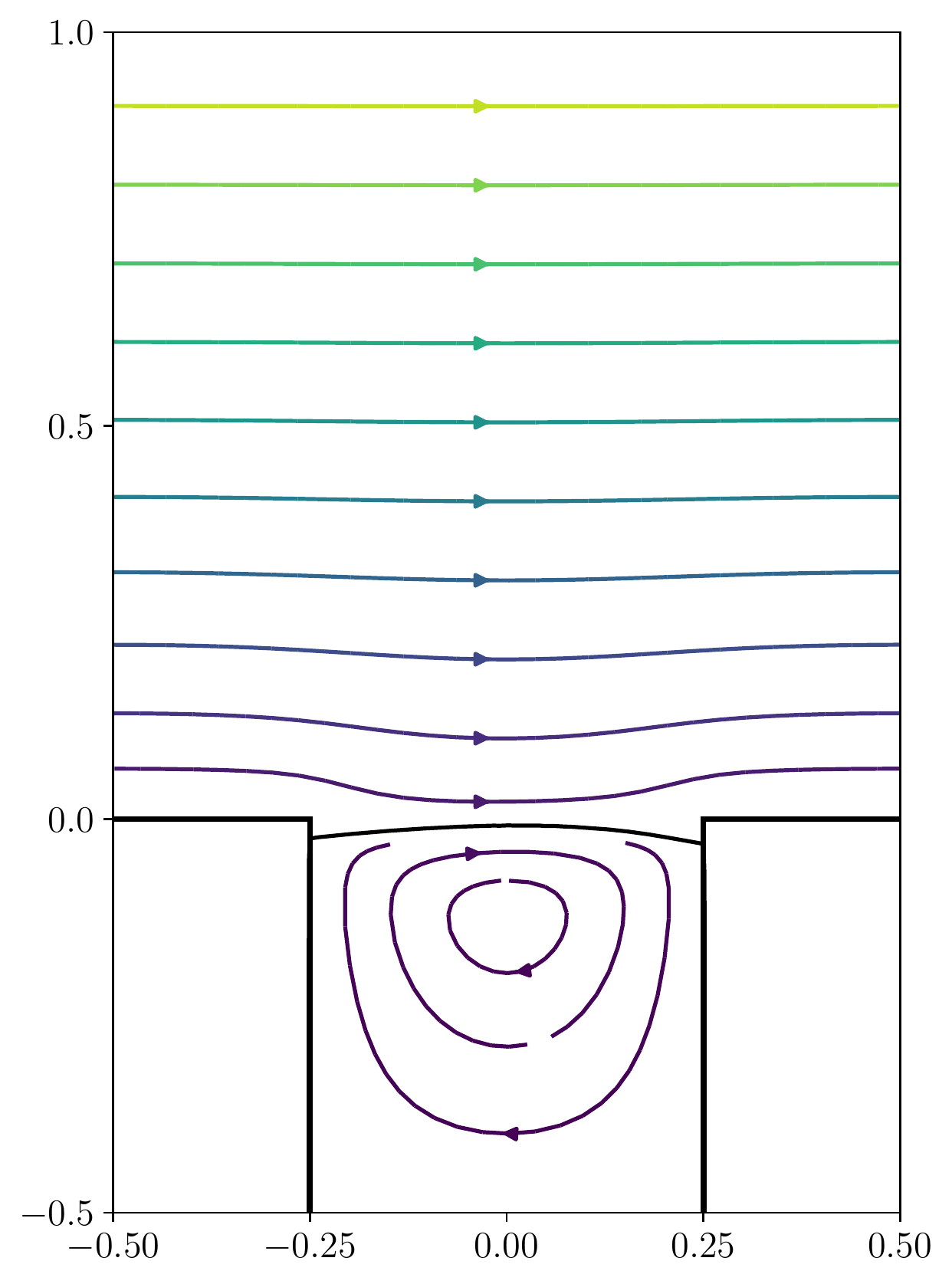}}
  \subfigure[$\quad \theta_s=105\degree$]
  {\includegraphics[width=.35\columnwidth]{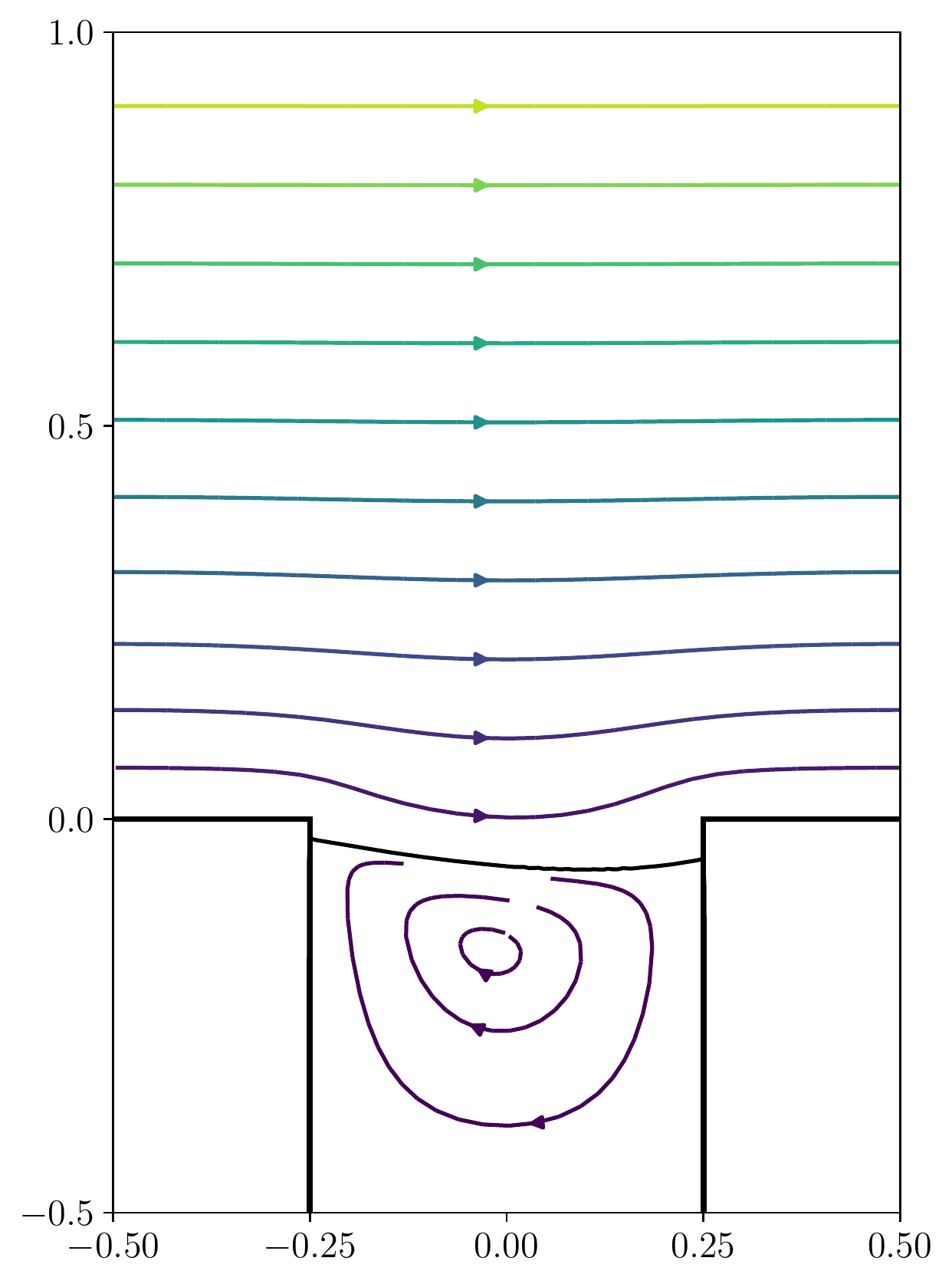}}


  \caption{(color online) Typical shapes of stable interfaces and streamlines for the flow over a partially filled cavity. These examples correspond to the steady state configurations, for (a) $\theta_s=80\degree$ and (b) $\theta_s=105\degree$, at $\tilde{\mu}_2/\tilde{\mu}_1=0.5$ and Ca $=0.02$.}
  \label{fig: shape}
\end{figure}

First, we examine typical interface profiles, both convex and concave, under moderate shear levels, see Fig.\ \ref{fig: shape}. Specifically, we consider the viscosity ratio $\tilde{\mu}_2/\tilde{\mu}_1=0.5$, the capillary number Ca $=0.02$, and the initial contact angle $\theta_0 = \theta_s$. The steady-state solutions are taken at $t=5$ in units of $1/\tilde{\dot{\gamma}}$.
\Ge{As illustrated in the figure, the flow, while circulating inside the cavity, is already parallel at $y \approx 0.5H$. The deformation of the interfaces is almost negligible comparing to the initial conditions, only the contact points displacing slightly in opposite directions due to the shear. These two configurations are examples of lubricant-infused cavities in working condition. The overall small change of the interface shapes is the reason for the weak shear dependence of the effective slip length discussed in Sec.\ \ref{eslip}.}

\begin{figure}[t]
 \begin{center}
 \subfigure[$\quad \theta_s=80 \degree$]
 {\includegraphics[width=.4\columnwidth]{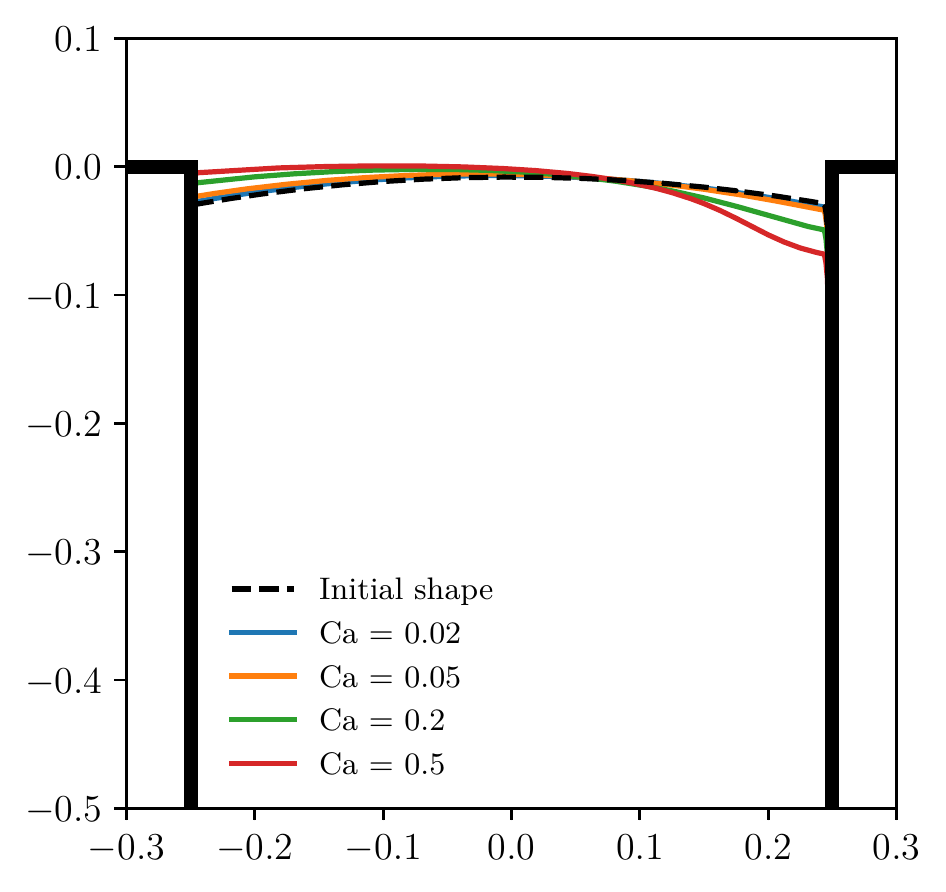}}
 \subfigure[$\quad \theta_s=105\degree$]
 {\includegraphics[width=.4\columnwidth]{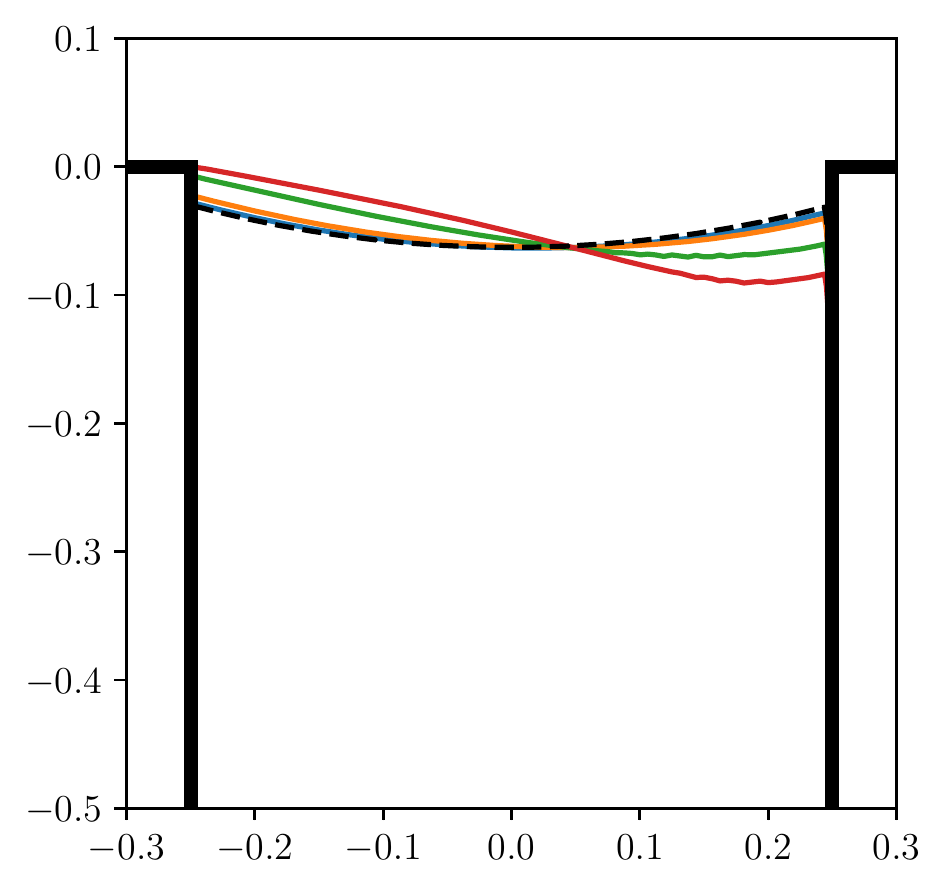}}
 \end{center}
 \caption{Interface profiles under increasing capillary numbers for viscosity ratio $\tilde{\mu}_2/\tilde{\mu}_1=5.05\e{-2}$ at $t=5$.}
 \label{fig: meniscus 2}
\end{figure}

\begin{figure}[t]
 \begin{center}
 \subfigure[$\quad \theta_s=80 \degree$]
 {\includegraphics[width=.4\columnwidth]{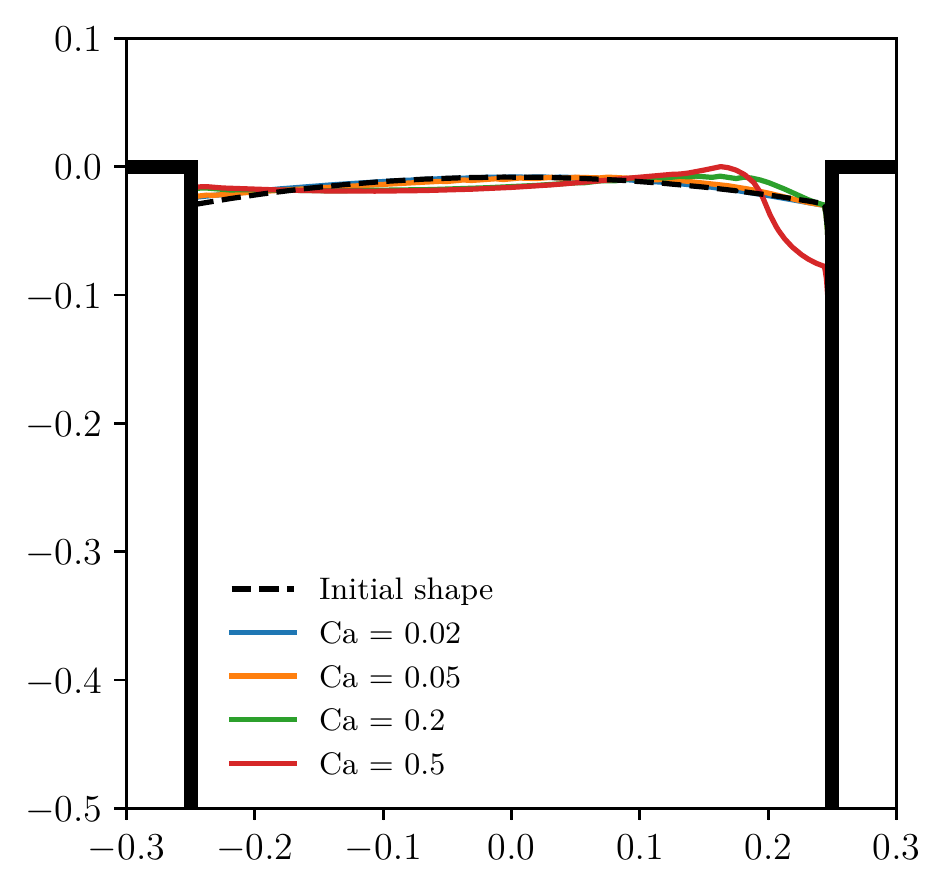}}
 \subfigure[$\quad \theta_s=105\degree$]
 {\includegraphics[width=.4\columnwidth]{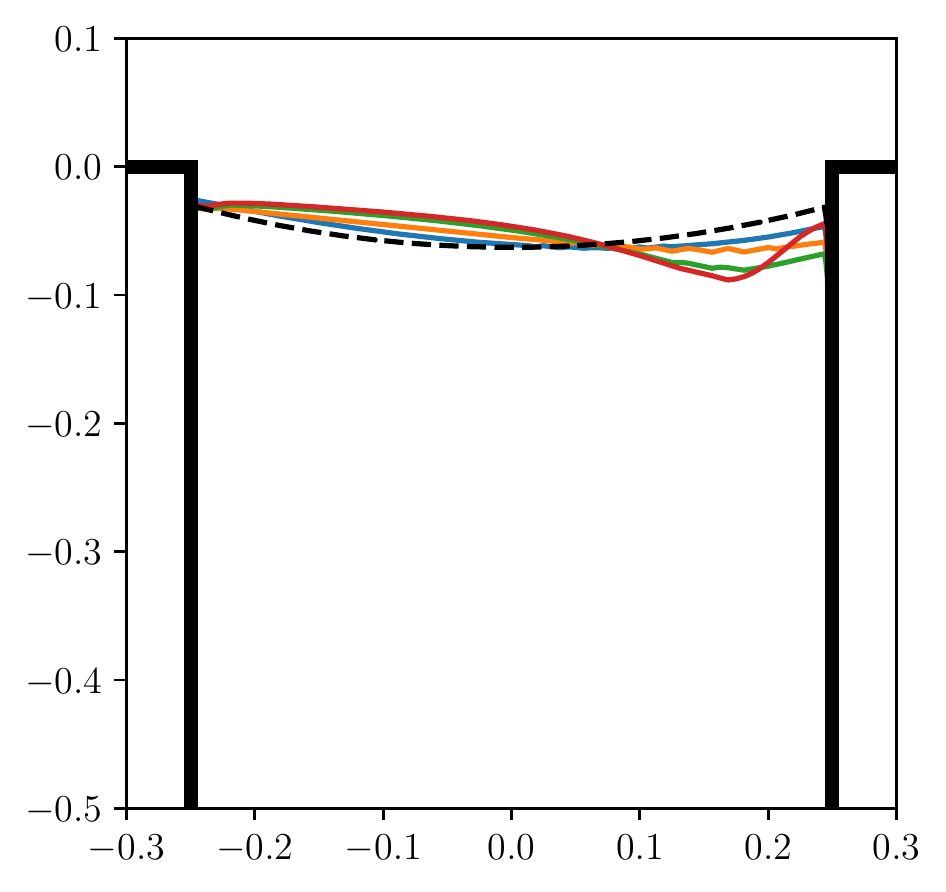}}
 \end{center}
 \caption{Interface profiles under increasing capillary numbers for viscosity ratio $\tilde{\mu}_2/\tilde{\mu}_1=0.5$ at $t=5$.}
 \label{fig: meniscus 2a}
\end{figure}

%

\begin{figure}[t]
 \begin{center}
 \subfigure[$\quad \theta_s=80 \degree$]
 {\includegraphics[width=.4\columnwidth]{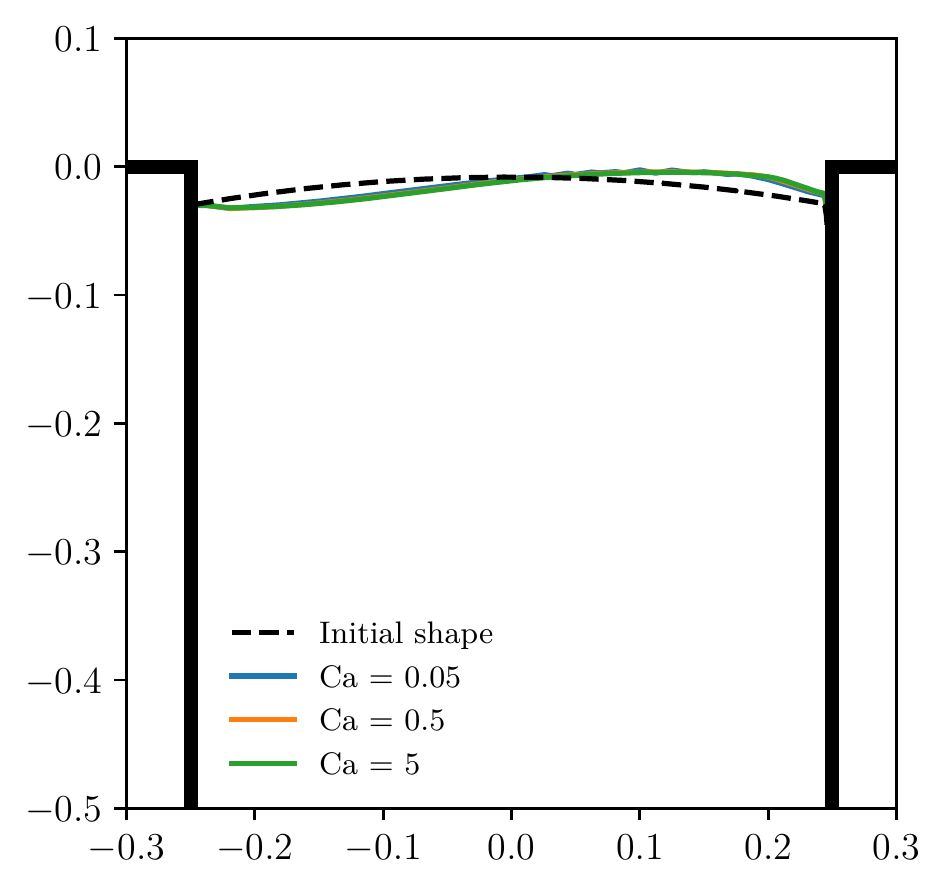}}
 \subfigure[$\quad \theta_s=105\degree$]
 {\includegraphics[width=.4\columnwidth]{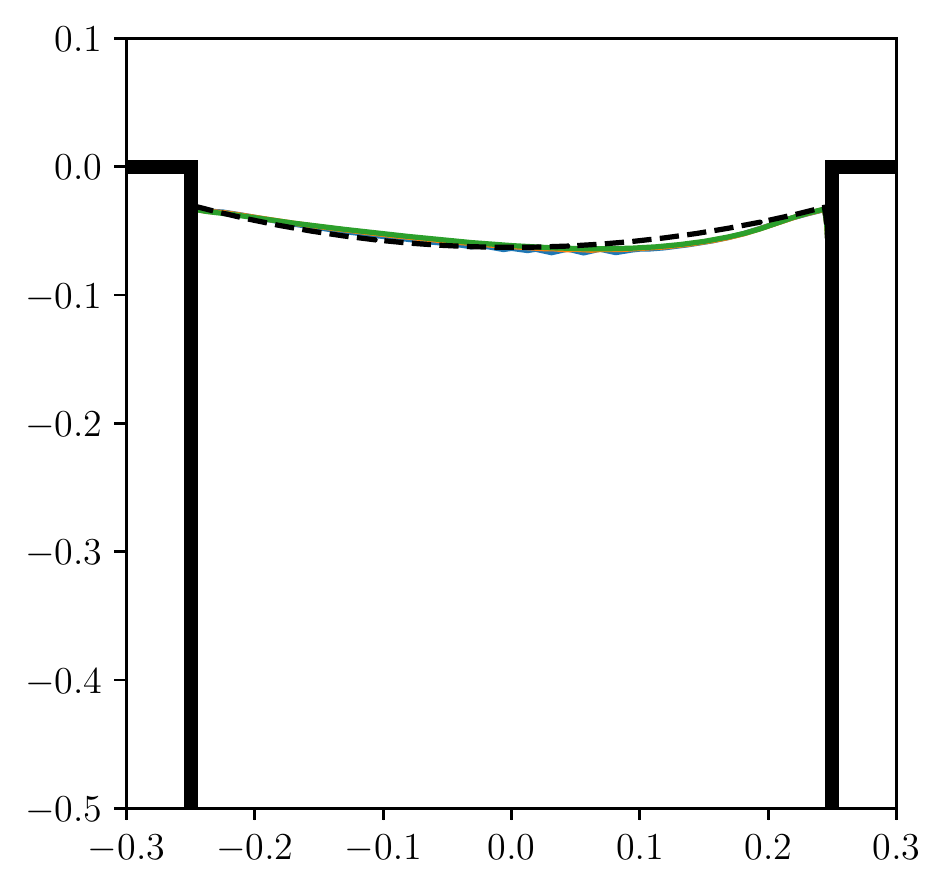}}
 \end{center}
 \caption{Interface profiles under increasing capillary numbers for viscosity ratio $\tilde{\mu}_2/\tilde{\mu}_1=3.17$ at $t=5$.}
 \label{fig: meniscus 1}
\end{figure}

\Ge{To test the robustness of the LIS under stronger shear, we successively increase the capillary number from 0.02 to 5, keeping the other parameters unchanged (see Tab.\ \ref{tab: param}). The resulting interface profiles are visualized in Figs.\ \ref{fig: meniscus 2}--\ref{fig: meniscus 1}.
As expected, increasing Ca generally leads to larger deformations of the interface. For $\tilde{\mu}_2/\tilde{\mu}_1=5.05\e{-2}$ (Fig.\ \ref{fig: meniscus 2}), the upstream contact point continuously moves towards the tip of the cavity, indicating a draining motion of the lubricant driven by the shear. For $\tilde{\mu}_2/\tilde{\mu}_1=0.5$ (Fig.\ \ref{fig: meniscus 2a}), the downstream contact point responds more instead, almost leading to the interface rupture when Ca $=0.5$.
However, as we further increase the viscosity ratio, increasing the shear has no visible effect on the interface. 
For $\tilde{\mu}_2/\tilde{\mu}_1=3.17$ (Fig.\ \ref{fig: meniscus 1}), we barely observe any additional deformation even increasing Ca by two orders of magnitude. The lubricant stays firmly in the cavity regardless of the external shear.}


\begin{figure}[t]
 \begin{center}
 \subfigure[$\quad \theta_s=80 \degree$]{\includegraphics[width=.49\columnwidth]{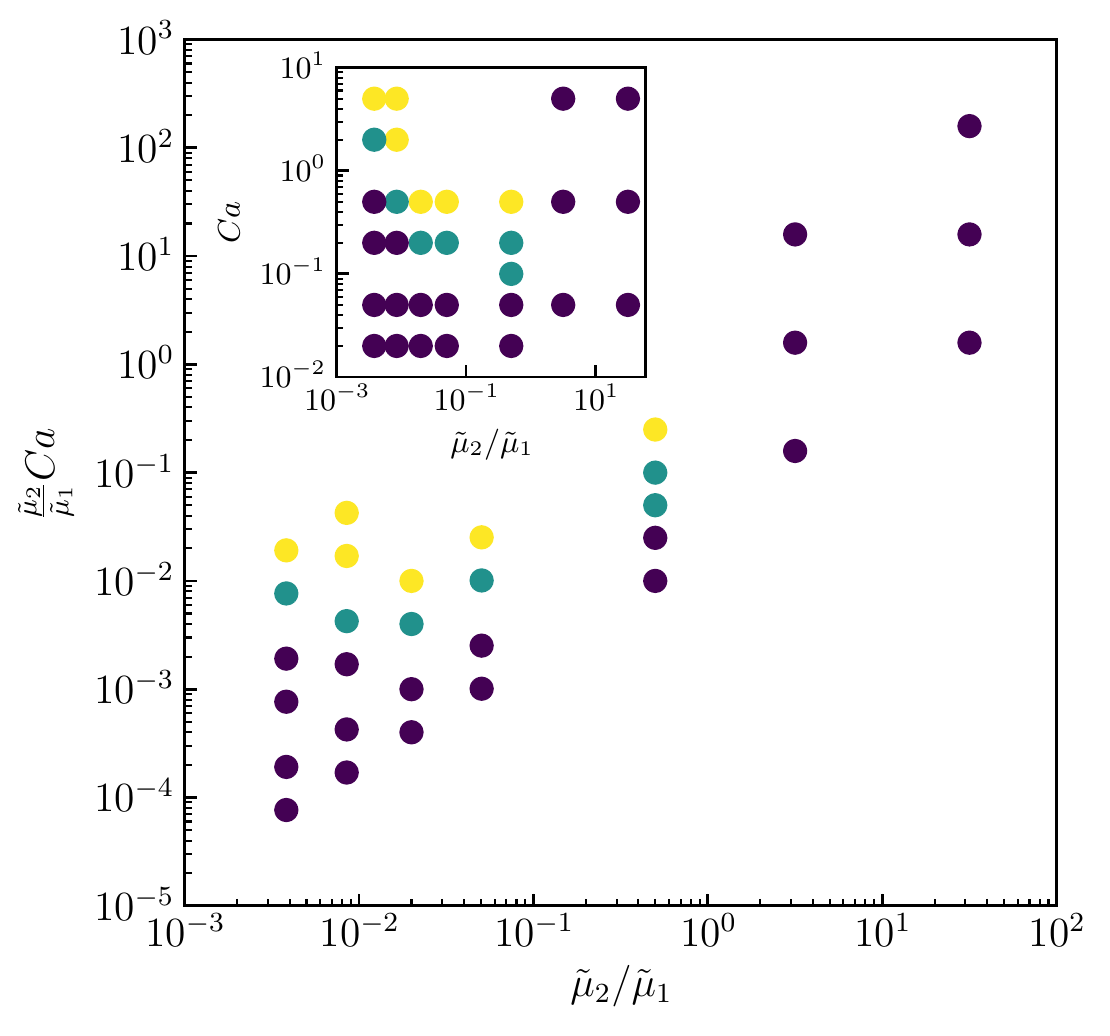}}
 \subfigure[$\quad \theta_s=105\degree$]{\includegraphics[width=.49\columnwidth]{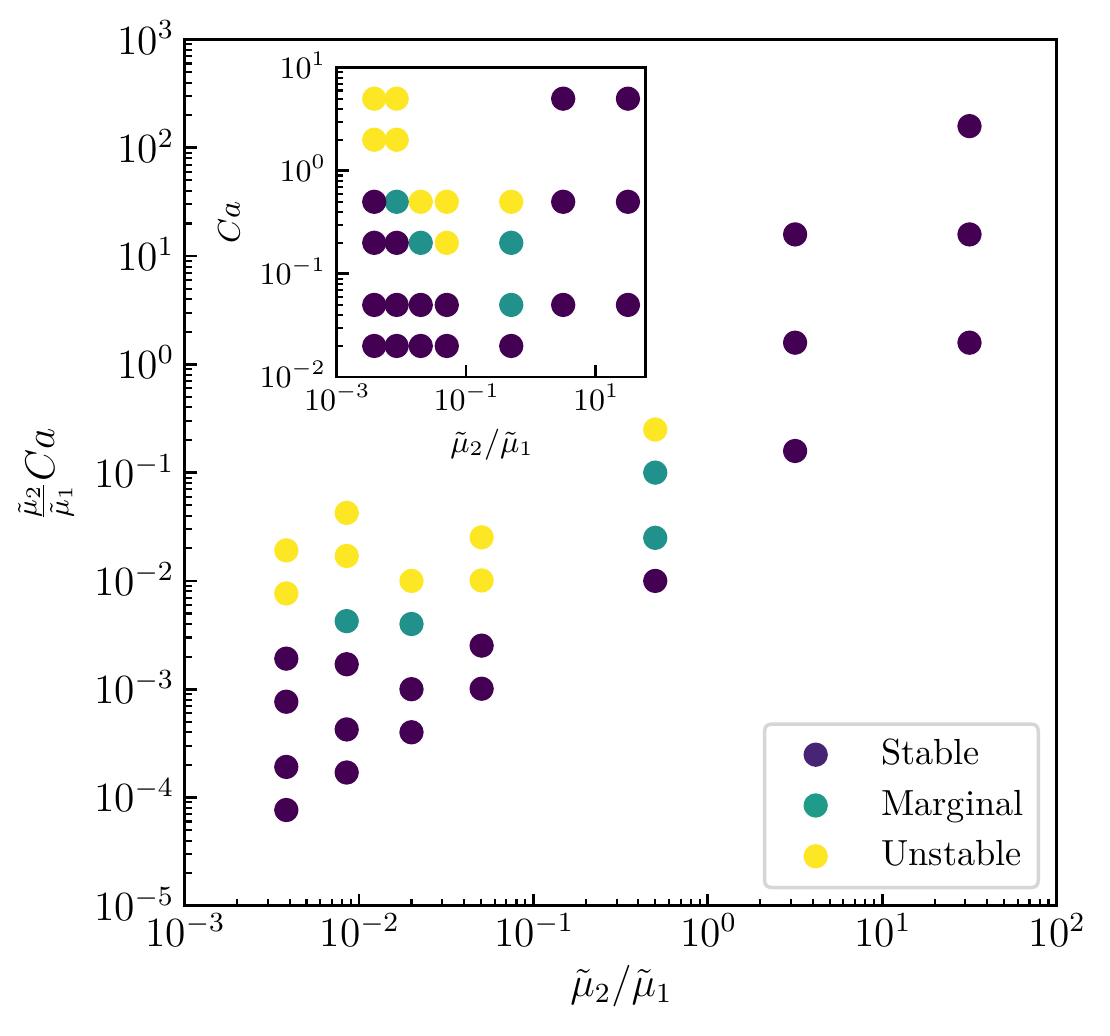}}
 \end{center}
 \begin{picture}(0,0)
   \put(-55,75){\includegraphics[height=1.8cm]{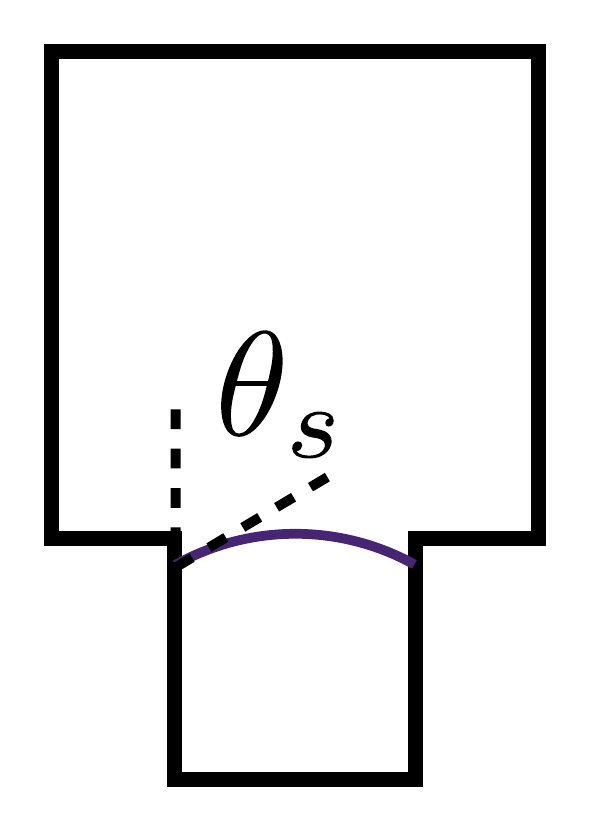}}
 \end{picture}
 \caption{Phase diagram in the cavity capillary--viscosity ratio plane ($\tilde{\mu}_2/\tilde{\mu}_1$, $\tilde{\mu}_2/\tilde{\mu}_1$Ca) showing the robustness of the lubricant-infused cavities under various capillary numbers and viscosity ratios. 
 The inset reports the same data as function of the outer capillary number Ca.}
 \label{fig: drain}
\end{figure}

\Ge{Following these observations, we map the results pertaining all the viscosity ratios and capillary numbers considered in the phase diagram in Fig.\ \ref{fig: drain}. Here, three regimes  are defined, which we label as stable, marginal, and unstable. For partially filled cavities with initial filling fraction $\delta=0.94$ (\ie initial depth $d_0/H=94\%$ measured from the contact points), we consider cases where the final depth varies within $94 \pm 2\%$ as \textit{stable}; if the final depth varies between $94 \pm 4\%$, which is very close to the cavity tip but still below, we consider the configuration  as \textit{marginal}; lastly, if one contact point has already/nearly hit the cavity tip, or if the interface is clearly disrupted (see \eg Fig.\ \ref{fig: meniscus 2a}), we consider the case as \textit{unstable}. A similar, but simplified, criterion has also been chosen in \cite{Seo_etal_18} for the onset of gas pocket instability in turbulent flows. We evaluate the final depths either in the steady states, or at $t=5$ if a steady state has not been reached.}


\Ge{As shown in Fig.\ \ref{fig: drain}, the robustness of the LIS exhibits a rather complex dependence on the capillary number and the viscosity ratio.
On the lower viscosity side, \ie $\tilde{\mu}_2/\tilde{\mu}_1 \lesssim 0.1$, lubricants of both convex and concave interfaces becomes unstable above a critical capillary number. Mapping the data on the ($\tilde{\mu}_2/\tilde{\mu}_1$, $\tilde{\mu}_2/\tilde{\mu}_1$Ca) plane, our results suggest Ca$_{crit} \approx 0.01 \tilde{\mu}_1/\tilde{\mu}_2$. That is, the critical capillary number, defined with the outer fluid, is inversely proportional to the viscosity ratio $\tilde{\mu}_2/\tilde{\mu}_1$; it is harder to drain a less viscous lubricant outside the cavity fixing the outer fluid. Note that similar results were also observed experimentally for longitudinal grooves, where less viscous lubricants are found to remain over a longer distance within the grooves \cite{Liu}. Our simulations thus point towards the same direction in the design of transverse LIS against shear-driven failures.}
 
\Ge{As the viscosity ratio further increases, however, the existence of a critical capillary number, Ca$_{crit}$, is no longer clear: at $\tilde{\mu}_2/\tilde{\mu}_1 =0.5$, the equivalent Ca$_{crit} \approx 0.1 \tilde{\mu}_1/\tilde{\mu}_2$, which would be an order-of-magnitude higher than before; while for $\tilde{\mu}_2/\tilde{\mu}_1 > 1$, no Ca$_{crit}$ has been found within a reasonable range of capillary numbers.
This shear-induced failure can be associated with draining in the cavities, and we propose that its mechanism be linked to the dewetting of the lubricant and to its viscosity.}

\Ge{Recalling the $\chi u_c ( \theta)$ relations for various viscosity ratios in Fig.\ \ref{fig:tabulated}, the profiles collapse onto one curve for $\tilde{\mu}_2/\tilde{\mu}_1 < 0.05$, suggesting that it is the cavity capillary number, $\tilde{\mu}_2/\tilde{\mu}_1$Ca, that determines the onset of failure. This is confirmed by testing one additional case with $\tilde{\mu}_2/\tilde{\mu}_1 = 0.02$ and the same dependency  $\chi u_c ( \theta)$ as in the other cases (thus not extracted from the phase-field simulations); for this case, we indeed obtain the same Ca$_{crit} \approx 0.01 \tilde{\mu}_1/\tilde{\mu}_2$, consistent with the other viscosity ratios in the same range (see Fig.\ \ref{fig: drain}).
When $\tilde{\mu}_2/\tilde{\mu}_1$ increases to above 0.5, Fig.\ \ref{fig:tabulated} shows a continuous deviation of the wetting relations. The slope of the $\chi u_c (\theta)$ curve near the static angle $\theta_s$ increases (in magnitude) rapidly with $\tilde{\mu}_2/\tilde{\mu}_1$, making it more difficult for the contact line to deform, hence reducing the drainage of  the lubricant towards the cavity corner. Since the capillary number is limited by the shear rates and proportional to the scale of the micron-scale texture, substrates impregnated by very viscous lubricants are in practice very difficult to fail.}

\Ge{The above phenomenological mechanism suggests that, taking the lubricant viscosity as a design parameter, the intermediate viscosity ratios (\eg $\tilde{\mu}_2/\tilde{\mu}_1 = 0.01 \sim 1$ depending on the specific condition) are to be avoided in the application of LIS. This is consistent with previous experiments of longitudinal grooves towards the lower viscosity branch \cite{Liu}; 
more viscous lubricants, on the other hand,  seem to ensure higher robustness.}



\Ge{Finally, we note that the critical capillary numbers reported here should be considered as an estimate, since, in practice, draining of the lubricant will also depend on the physical/chemical conditions near the cavity corner. However, we do not expect these practical limitations to influence the qualitative insight obtained from our simulations.}

\section{Conclusions}

In this paper, motivated by applications of micro-engineered liquid-infused surfaces, we study the drag reduction and robustness of the flow over an array of two-dimensional tranverse grooves partially filled with an immiscible lubricant.

We use a multiscale numerical framework to model the wetting of the two fluids at the cavity walls as well as the deformation of the interface under the shear. In particular, we combine two separate simulation methods at different scales: \textit{(i)} nanoscale phase field simulations for the contact line dynamics, and \textit{(ii)} micron-scale Stokes flows simulations using information from \textit{(i)} as a modified boundary condition, assuming self-similarity of the velocity field in the vicinity of the moving contact line. We believe the approach is however more general and it could be extended to include molecular dynamics simulations modelling the surface chemistry and roughness at the nanoscale.

We examine the effective slip $\lambda_e$ in order to quantify the steady-state drag reduction of the LIS. Specifically, we fix the geometry of the cavity and vary the lubricant-to-outer-fluid viscosity ratio $\tilde{\mu}_2/\tilde{\mu}_1$, the capillary number Ca, the static contact angle $\theta_s$, and the filling fraction of the cavity $\delta$. 
\Ge{The main results are summarized as follows.
\begin{enumerate}
 \item \Ge{$\lambda_e$ depends primarily on $\delta$; the filling rate is therefore the main factor determining the effective slip.}
 \item \Ge{Lower $\tilde{\mu}_2/\tilde{\mu}_1$ leads to reduced drag; the reduction is however less pronounced comparing to fully covered cavities. 
 We relate this effect to the shear stress profiles $\tau_{xy}$ along the cavity tip, and show that $\tau_{xy}$ is non-uniform (contrary to the fully covered cases).}
 \item \Ge{The effect of the contact angle on the effective slip length is different for different viscosity ratios and the filling fractions.}
 \item \Ge{The effect of the contact angle hysteresis and of the capillary number on $\lambda_e$ is negligible, except}
 \item \Ge{when Ca increases above a critical value Ca$_{crit}$, and the LIS can possibly fail. For an initial filling fraction $\delta =0.94$, the critical capillary number Ca$_{crit} \approx 0.01 \tilde{\mu}_1/\tilde{\mu}_2$ for $\tilde{\mu}_2/\tilde{\mu}_1 \lesssim 0.1$. For very viscous lubricants (\eg $\tilde{\mu}_2/\tilde{\mu}_1 >1$), on the other hand, the cavity remains impregnated due to their generally larger contact line velocity.}
\end{enumerate}}


As a final remark, we note that this problem is characterised by a large number of control parameters, including \eg geometry, static contact angle, surface chemistry, so that this study can be extended in a number of non-trivial ways. In addition, from a purely hydrodynamic point of view, the flow above the cavity may affect the contact line motion: it may therefore be relevant to study the response of the flow in the cavity to temporally varying shear and vortices relatively far from it.

\section*{Acknowledgments}

We thank Shervin Bagheri and U\'{g}is L\={a}cis for calling our attention on the problem of LIS. We also thank Mauro Chinappi and Pengyu Lv for many helpful discussions. This project is funded by the European Union Horizon 2020 research and innovation programme under Grant Agreement No.\ 664823 and the Swedish Research Council (No.\ 621-2012-2360).

{}




\begin{thebibliography}{99}


\bibitem{Watanabe} K. Watanabe, Yanuar, and H. Udagawa, 
Drag reduction of Newtonian fluid in a circular pipe with a highly water-repellent wall,
J. Fluid Mech. \textbf{381}, 225--238 (1999).

\bibitem{Ou} J. Ou, B. Perot, and J. P. Rothstein, 
Laminar drag reduction in microchannels using ultrahydrophobic surfaces,
Phys. Fluids \textbf{16}, 4635 (2004).

\bibitem{Schaffel} D. Sch\"{a}ffel, K. Koynov, D. Vollmer, H.-J. Butt, and C. Sch\"{o}necker, 
Local flow field and slip length of superhydrophobic surfaces,
Phys. Rev. Lett. \textbf{116}, 134501 (2016).

\bibitem{Lee}  C. Lee, C.-H. Choi, and C.-J. Kim, 
Superhydrophobic drag reduction in laminar flows: a critical review,
Exp. Fluids \textbf{57}, 176 (2016).

\bibitem{Choi_Kim}  C.-H. Choi, and C.-J. Kim, 
Large slip of aqueous liquid flow over a nanoengineered superhydrophobic surface,
Phys. Rev. Lett. \textbf{96}, 066001 (2006).

\bibitem{Steinberger}  A. Steinberger, C. Cottin-Bizonne, P. Kleimann, and E. Charlaix, 
High friction on a bubble mattress,
Nat. Mater. \textbf{6}, 665 (2007).

\bibitem{Karatay}  E. Karatay, A.S. Haase, C.W. Visser, C. Sun, D. Lohse, P.A. Tsai, R.G.H.,
Control of slippage with tunable bubble mattresses,
Lammertink, Proc. Natl. Acad. Sci. \textbf{110}, 8422-8426 (2013).

\bibitem{Bocquet} L. Bocquet, and E. Lauga, 
A smooth future?,
Nat. Mater. \textbf{10}, 334 (2011).

\bibitem{Gentili} D. Gentili, G. Bolognesi, A. Giacomello, M. Chinappi, and C. M. Casciola, 
Pressure effects on water slippage over silane-coated rough surfaces: pillars and holes,
Microfluid Nanofluid \textbf{16}6, 1009--1018 (2014).

\bibitem{Giacomello} A. Giacomello, M. Chinappi, S. Meloni, and C. M. Casciola, 
Metastable Wetting on Superhydrophobic Surfaces: Continuum and Atomistic Views of the Cassie-Baxter–Wenzel Transition,
Phys. Rev. Lett. \textbf{109}, 226102 (2012).

\bibitem{Solomon} B. R. Solomon, K. S. Khalil, and K. K. Varanasi, 
Drag reduction using lubricant-impregnated surfaces in viscous laminar flow,
Langmuir \textbf{30}, 10970--10976 (2014).

\bibitem{Rosenberg} B. J. Rosenberg, T. V. Buren, M. K. Fu, and A. J. Smits, 
Turbulent drag reduction over air- and liquid- impregnated surfaces,
Phys. Fluids \textbf{28}, 015103 (2016).

\bibitem{Hemeda} A. A. Hemeda, and H. Vahedi Tafreshi, 
Liquid–infused surfaces with trapped air (LISTA) for drag force reduction,
Langmuir \textbf{32}, 2955--2962 (2016).

\bibitem{Wexler}  J. S. Wexler, I. Jacobi, and H. A. Stone, 
Shear-driven failure of liquid-infused surfaces,
Phys. Rev. Lett. \textbf{114}, 168301 (2015).

\bibitem{Jacobi}  I. Jacobi, J. S. Wexler, and H. A. Stone, 
Overflow cascades in liquid-infused substrates,
Phys. Fluids \textbf{27}, 082101 (2015).

\bibitem{Liu}  Y. Liu, J. S. Wexler, C. Sch\"{o}necker, and H. A. Stone, 
Effect of viscosity ratio on the shear-driven failure of liquid-infused surfaces,
Phys. Rev. Fluids \textbf{1}, 074003 (2016).


\bibitem{Lauga_Stone} E. Lauga, and H. A. Stone, 
Effective slip in pressure-driven Stokes flow,
J. Fluid Mech. \textbf{489}, 55--77 (2003).

\bibitem{Sbragalia_Prosperetti} M. Sbragalia, and A. Prosperetti, 
A note on the effective slip properties for microchannel flows with ultrahydrophobic surfaces,
Phys. Fluids \textbf{19}, 043603 (2007).

\bibitem{Davis_Lauga} A. M. J. Davis, and E. Lauga, 
Geometric transition in friction for flow over a bubble mattress,
Phys. Fluids \textbf{21}, 011701 (2009).

\bibitem{Ng_Wang} C.-O. Ng, and C. Y. Wang, 
Stokes shear flow over a grating: implications for superhydrophobic slip,
Phys. Fluids \textbf{21}, 013602 (2009).


\bibitem{Schonecker} C. Sch\"{o}necker, T. Baier, and S. Hardt, 
Influence of the enclosed fluid on the flow over a microstructured surface in the Cassie state,
J. Fluid Mech. \textbf{740}, 168--195 (2014).

\bibitem{Nizkaya} T. V. Nizkaya, E. S. Asmolov, and O. I.  Vinogradova, 
Gas cushion model and hydrodynamic boundary conditions for superhydrophobic textures,
Phys. Rev. E \textbf{90}, 043017 (2014).

\bibitem{Crowdy_long} D. G. Crowdy, 
Perturbation analysis of subphase gas and meniscus curvature effects for longitudinal flows over superhydrophobic surfaces,
J. Fluid Mech. \textbf{822}, 307--326 (2017).

\bibitem{Crowdy_tran} D. G. Crowdy,
Slip length for transverse shear flow over a periodic array of weakly curved menisci,
Phys. Fluids \textbf{29}, 9 (2017).  

\bibitem{Davies_etal} J. Davies, D. Maynes, B. W. Webb, and B. Woolford,
Laminar flow in a microchannel with superhydrophobic walls exhibiting transverse ribs,
Phys. Fluids \textbf{18}, 087110 (2006).

\bibitem{Martell_Perot_Rothstein} M. B. Martell, J. B. Perot, J. P. Rothstein, 
Direct numerical simulations of turbulent flows over superhydrophobic surfaces,
J. Fluid Mech. \textbf{620}, 31--41 (2009).

\bibitem{Cheng_Teo_Khoo} Y. P. Cheng, C. J. Teo, and B. C. Khoo,
Microchannel flows with superhydrophobic surfaces: effects of Reynolds number and pattern width to channel height ratio,
Phys. Fluids \textbf{21}, 122004 (2009).

\bibitem{Wang_Teo_Khoo} L. P. Wang, C. J. Teo, and B. C. Khoo,
Effects of interface deformation on flow through microtubes containing superhydrophobic surfaces with longitudinal ribs and grooves,
Microfluid Nanofluid 16:225--236 (2014).

\bibitem{Teo_Khoo_14} C. J. Teo, and B. C. Khoo,
Effects of interface curvature on Poiseuille flow through microchannels and microtubes containing superhydrophobic surfaces with transverse grooves and ribs,
Microfluid Nanofluid 17:891--905 (2014).

\bibitem{Seo_etal_18} J. Seo, R. Garcia-Mayoral, A. Mani,
Turbulent flows over superhydrophobic surfaces: flow-induced capillary waves, and robustness of air-water interfaces,
J. Fluid Mech. \textbf{835}, 45--85 (2018).

\bibitem{finland}  J. Hyv\"{a}luoma, and J. Harting, 
Slip flow over structured surfaces with entrapped microbubbles,
Phys. Rev. Lett. \textbf{100}, 246001 (2008).

\bibitem{finland2}  J. Hyv\"{a}luoma, C. Kunert, J. Harting, 
Simulations of slip flow on nanobubble-laden surfaces,
J. Phys.: Condens. Matter. \textbf{23}, 184106 (2011).

\bibitem{Gao_Peng} P. Gao, and J. J. Feng, 
Enhanced slip on a patterned substrate due to depinning of contact line,
Phys. Fluids \textbf{21}, 102102 (2009).



\bibitem{Batchelor} G. K. Batchelor, {\it An introduction to fluid dynamics}, (Cambridge university press, 2000).

\bibitem{Huh_Scriven}  C. Huh and L. E. Scriven, 
Hydrodynamic model of steady movement of a solid/liquid/fluid contact line,
J. Colloid Int. Sci. \textbf{35}, No.\ 1 (1971).

\bibitem{Jacqmin2000} D. Jacqmin,
Contact-line dynamics of a diffuse fluid interface,
 J. Fluid Mech. \textbf{402}, 57--88 (2000).

\bibitem{Johansson} P. Johansson, A. Carlson, and B. Hess, 
Water--substrate physico-chemistry in wetting dynamics,
J. Fluid Mech. \textbf{781}, 695--711 (2015).

\bibitem{Sbragaglia_etal} M. Sbragaglia, R. Benzi, L. Biferale, S. Succi, F. Toschi,
Surface roughness-hydrophobicity coupling in microchannel and nanochannel flows,
Phys. Rev. Lett. \textbf{97}, 204503 (2006).

\bibitem{COX} R. G. Cox,
The dynamics of the spreading of liquids on a solid surface. Part 1. Viscous flow,
J. Fluid Mech. \textbf{168}, 169--194 (1986).

\bibitem{Martin}  M. Kronbichler and G. Kreiss,
A phase-field microscale enhancement for macro models of capillary-driven contact point dynamics,
J. Comp. Multiphase Flows. 9(3) 114--126 (2017).

\bibitem{Hanna}  H. Holmgren and G. Kreiss,
A Computational Multiscale Model for Contact Line Dynamics,
ArXiv 1709.04917 (2017).




\bibitem{MartinHPC} M. Kronbichler, A. Diagne and H. Holmgren,
A fast massively parallel two-phase flow solver for microfluidic chip simulation,
 Int. J. High Perform. Comput. Appl. \textbf{}, 1--22 (2016).

\bibitem{CONSLS} E. Olsson and G. Kreiss,
A conservative level set method for two phase flow,
 J. Comput. Phys. \textbf{210}, 225 - 246 (2005).


\bibitem{DEAL1} W. Bangerth, R. Hartmann and G. Kanschat,
deal. II -- a general-purpose object-oriented finite element library,
 ACM Trans. Math. Softw. \textbf{33}, (2007).

\bibitem{DEAL2} W. Bangerth and D. Davydov and T. Heister and L. Heltai and G. Kanschat and M. Kronbichler and M. Maier and B. Turcksin and D. Wells, 
The \texttt{deal.II} Library, Version 8.4,
J. Numer. Math. \textbf{24} 3, 135--141 (2016).




\bibitem{dimension} We estimate the dimensional shear rate at the upper boundary of our computational domian from the experimental Hele-Shaw setup in \cite{Wexler} (2 mL/min through a cross-section of 180 $\mu$m $\times$ 7 mm). Assuming a parabolic velocity profile, the reference velocity at $\tilde{H}=20$ $\mu$m is $\tilde{U}=$ 0.016 m/s, leading to a shear rate of $\dot{\tilde{\gamma}}=800$ s$^{-1}$. The characteristic contact line velocity is calculated as $\tilde{U}_c=\tilde{\sigma}/\tilde{\mu}_2$, taking $\tilde{\sigma}=0.02$ kg/$s^2$ and $\tilde{\mu}_2=0.0076$ $\nicefrac{kg}{ms}$, as in \cite{Martin}.


\end{thebibliography}
\end{document}